\documentclass[12pt]{article}

\usepackage{amsbsy, amstext, amssymb, amsthm, amsmath,bm,wasysym}
\usepackage{algorithmic,graphicx,booktabs,hyperref,rotating,enumitem}
\usepackage[margin=1in]{geometry}
\usepackage[lined,boxed,linesnumbered]{algorithm2e}
\usepackage{array,stmaryrd}
\usepackage{tabularx}
\usepackage{pdflscape}
\usepackage{adjustbox}
\usepackage{subfig}
\usepackage{longtable}
\usepackage{authblk}

\usepackage{subfiles}

\usepackage[natbibapa]{apacite}
\usepackage{etoolbox}
\patchcmd{\APACjournalVolNumPages}{\unskip({#3})}{}{}{} 
\patchcmd{\APACjournalVolNumPages}{\Bem{#2}}{#2}{}{}

\AtBeginEnvironment{APACrefURL}{\renewcommand{\url}[1]{}}
\AtBeginDocument{}

\usepackage{tikz}
\usetikzlibrary{shapes.geometric, arrows}
\usetikzlibrary{arrows.meta}
\usetikzlibrary{shapes,backgrounds,calc,positioning}

\usepackage{blkarray}

\newcommand{\diag}{\mathrm{diag}}

\newcommand{\Cbf}{{\bm C}}
\newcommand{\Dbf}{{\bm D}}

\newcommand{\Tbf}{{\bm T}}

\newcommand{\Xbf}{{\bm X}}
\newcommand{\Ybf}{{\bm Y}}
\newcommand{\Zbf}{{\bm Z}}

\newcommand{\bbf}{{\bm b}}

\usepackage{amsmath}
\usepackage{array}
\usepackage{chngcntr}
\counterwithout{table}{section}
\renewcommand{\thetable}{\arabic{table}}

\title{Efficient Implementation of a Semiparametric Joint Model for Multivariate Longitudinal Biomarkers and Competing Risks Time-to-Event Data}

\author[1,3]{Shanpeng Li*}
\author[2]{Emily Ouyang*}
\author[3]{Jin Zhou}
\author[2]{Xinping Cui}
\author[3]{Gang Li$^{+}$}

\affil[1]{Department of Computational and Quantitative Medicine, City of Hope, Duarte, CA, USA.}
\affil[2]{Department of Biostatistics, University of California at Riverside, Riverside, CA, USA.}
\affil[3]{Department of Biostatistics, University of California at Los Angeles, Los Angeles, CA, USA.}

\usepackage{xr}
\usepackage{comment}
\usepackage{color}
\makeatletter
\newcommand*{\addFileDependency}[1]{
  \typeout{(#1)}
  \@addtofilelist{#1}
  \IfFileExists{#1}{}{\typeout{No file #1.}}
}
\makeatother

\begin{document}

\date{}
\maketitle

\begin{abstract}
Joint modeling has become increasingly popular for characterizing the association between one or more longitudinal biomarkers and competing risks time-to-event outcomes. However, semiparametric multivariate joint modeling for large-scale data encounter substantial statistical and computational challenges, primarily due to the high dimensionality of random effects and the complexity of estimating nonparametric baseline hazards. These challenges often lead to prolonged computation time and excessive memory usage, limiting the utility of joint modeling for biobank-scale datasets. In this article, we introduce an efficient implementation of a semiparametric multivariate joint model, supported by a normal approximation and customized linear scan algorithms within an expectation-maximization (EM) framework. Our method significantly reduces computation time and memory consumption, enabling the analysis of data from thousands of subjects. The scalability and estimation accuracy of our approach are demonstrated through two simulation studies. We also present an application to the Primary Biliary Cirrhosis (PBC) dataset involving five longitudinal biomarkers as an illustrative example. A user-friendly R package, \texttt{FastJM}, has been developed for the shared random effects joint model with efficient implementation. The package is publicly available on the Comprehensive R Archive Network: \url{https://CRAN.R-project.org/package=FastJM}.
\end{abstract}


\noindent * These authors contributed equally to this work.

\noindent $^{+}$ Corresponding author: Gang Li vli@ucla.edu

\section{Introduction}
In clinical research and other longitudinal studies, it is common to collect both longitudinal and time-to-event data  on each participant, and these two outcomes are often correlated.
Joint models of longitudinal and survival data have been widely used to mitigate incorrect estimation and statistical inferences associated with separate analyses/two-stage modeling approach \citep{ElashoffLiLi17JMBook, Rizopoulos10JM}. For instance, in longitudinal data analysis, joint models are often used to handle {\it nonignorable} missing data due to a terminal event, which cannot be properly accounted for by a standard  mixed effects model or generalized estimating equation (GEE) method that relies on the {\it ignorable} missing-at-random or missing-completely-at-random assumption \citep{henderson2000joint, elashoff2008joint, sattar2019joint}. 
Joint models are also popularly employed in survival analysis to study the effects of a time-dependent covariate that is measured intermittently or subject to measurement error \citep{tsiatis2004joint,Wang06CorrectedScore, SongDavidianTsiatis02SemiparamJM, crowther2013joint}. Traditional joint models and their applications have predominantly focused on the form of a joint modeling specification with a single longitudinal outcome i.e., univariate joint modeling. More recently, joint models with multiple longitudinal outcomes (multivariate joint modeling) have been receiving lots of attentions in the statistical literature since fitting a joint model for each longitudinal outcome separately does not take into account of correlation among various longitudinal biomarkers and could lead to biased estimates in the survival sub-model, particularly when omitted biomarkers are associated with both the target longitudinal outcome and the time-to-event process \citep{hickey2016joint,murray2023fast}, and could yield subpar prediction of clinical outcomes \citep{hickey2018joinerml}. However, current estimation algorithms/software packages for multivariate joint modeling still remained underdeveloped, as discussed below.

Over the past two decades, both frequentist and Bayesian approaches have contributed to the methodological development and estimation algorithms in the context of multivariate joint modeling. As the first publicly available software package based on a frequentist framework, \citet{hickey2018joinerml} proposed a Monte Carlo Expectation-Maximization (MCEM) algorithm to address the challenge of high-dimensional integration over random effects, namely the \texttt{joineRML} R package, offering an alternative to the Gauss-Hermite quadrature method originally proposed by \citet{wulfsohn1997joint} in the joint modeling context. More recently, \citet{murray2022fast,murray2023fast} proposed a fast approximate EM algorithm by assuming that the distribution of subject-specific random effects conditional on the observed data is (multivariate) normally distributed. This approach reduces each required integral to a uniformly one-dimensional problem, regardless of the dimensionality of random effects, thereby substantially improving computational efficiency. Nevertheless, both estimation algorithms have notable limitations. First, neither is capable of accommodating competing risks data, which are commonly seen in clinical studies and are inappropriate to treat as independent censoring \citep{putter2007tutorial,hickey2016joint}. Additionally, both approaches face scalability challenges in large-sample settings, due to the high dimensionality of hyperparameters associated with nonparametric baseline hazard specifications \citep{li2022efficient}. 

In response to these limitations, Bayesian joint modeling has emerged as a promising direction to address both theoretical and computational challenges encountered in current frequentist approaches. \citet{rstanarm2024} and \citet{rizopoulos2024optimizing} offered fully Bayesian probabilistic inference using Markov Chain Monte Carlo (MCMC) algorithms and further made publicly available software packages, namely \texttt{rstanarm} and \texttt{JMbayes2}, respectively. While both approaches provide flexible modeling capabilities, \texttt{JMbayes2} notably supports competing risks and accommodates many longitudinal biomarkers, whereas \texttt{rstanarm} is currently limited to a maximum of three longitudinal outcomes. Nonetheless, both estimation procedures are not scalable to large sample size data since multiple MCMC chains are often needed to obtain stable posterior estimation, which can encounter slow convergence and eventually become computationally expensive. To alleviate these limitations, \citet{rustand2024fast} introduced a Bayesian integrated nested Laplace approximation (INLA) algorithm, implemented in the \texttt{INLAjoint} R package \citep{rustand2024joint}, as an efficient and accurate alternative to MCMC-based inference. The core algorithm proceeds in three key steps: (1) approximating the marginal posterior distribution of the hyperparameters using a Laplace approximation; (2) approximating the conditional posterior distribution of the latent field given the hyperparameters; and (3) employing numerical integration to obtain the marginal posterior distributions of the latent field. To further improve computational speed, an empirical Bayes approach may be applied at the first step by using only the mode—rather than the full curvature—thus relaxing the computational burden caused by the fully Bayesian inference procedure. Despite its appealing computational efficiency, the INLA-based algorithm may struggle to accommodate multiple highly correlated longitudinal biomarkers in large-sample settings. This limitation may stem from several factors discussed in their work: (1) reliance on large sparse matrix computations for analytical posterior approximations; (2) substantial memory requirements for complex joint models; and (3) the original algorithmic design does not intend to deal with large number of hyperparameters, which becomes restrictive under the first two conditions.

The main contribution of this article is to offer an efficient implementation of a semiparametric joint model for multiple longitudinal biomarkers and a competing risks time-to-event outcome, with innovations in two key aspects:

\begin{enumerate}
    \item Instead of relying on Gauss-Hermite quadrature for numerical integration over high-dimensional random effects, we adopt a normal approximation approach introduced by \citet{murray2022fast,murray2023fast} within an EM framework. In contrast to \citet{murray2022fast,murray2023fast}, our estimation algorithms eliminate the need for one-dimensional numerical integration in the E-step by approximating the expectations of the necessary terms with analytical forms using the properties of the moment-generating function of the (multivariate) normal distribution. Our simulation studies demonstrate that our approximation algorithms yield satisfactory estimates.
    \item We incorporate the customized linear scan algorithms proposed by \citet{li2022efficient} into the multivariate joint modeling context. This allows us to reduce the computational complexity associated with estimating nonparametric baseline hazards—from non-linear growth to $O(n)$ operations—extending the efficiency gains shown in the univariate setting by \citet{li2022efficient} to multivariate joint models.
\end{enumerate}

The rest of our article is organized as follows. In Section \ref{Sec:methods}, we describe the multivariate joint model formulation, likelihood, estimation methods for parameters and standard errors, and the details of our efficient implementation. In Section \ref{Sec:simulations}, we present the finite sample properties of our estimation procedure. In Section \ref{Sec:application}, we illustrate the application of our joint model to a Primary biliary cirrhosis (PBC) study before concluding with discussions in Section \ref{Sec:discussion}.

\section{Methods}
\label{Sec:methods}
\subsection{Notations}
For each subject $i=1,\ldots,n$, we observe $\Ybf_i = (\Ybf_{i1}^{\top}, \ldots, \Ybf_{iG}^{\top})^{\top}$ where each $\Ybf_{ig}$ denotes the $g^{th}$ longitudinal response vector of interest, for $g=1,\ldots,G$. The $g^{th}$ longitudinal response for subject $i$, $\Ybf_{ig} = \left\{Y_{ig}(t_{i1g}), \ldots,Y_{ig}(t_{in_ig})\right\}$ is measured at time $t_{ijg}, j=1,\ldots,n_{ig}$, where $n_{ig}$ can differ across subjects and responses. In addition, each subject may experience one of $K (K\ge 1)$ distinct failure types or be right censored during the follow-up.  Let $\Tilde{T}_i$ denote the failure time of interest, $\Tilde{D}_i$ the failure type taking values in $\{1, \ldots, K\}$, and $C_i$ be an non-informative, independent censoring time for subject $i$. Then the observed right-censored competing risks time-to-event data for subject $i$ has the form  $(T_i, D_i)\equiv \left\{\text{min}(\Tilde{T}_i, C_i),\Tilde{D}_i I(\Tilde{T}_i \le C_i)\right\}$, $i=1,\ldots, n$. The observed data for all subjects are then denoted as $(\Ybf, \Tbf, \Dbf)$, with $\Ybf = \left\{\Ybf_1, \ldots, \Ybf_n\right\}$, $\Tbf = \left\{T_1, \ldots, T_n\right\}$, and $\Dbf = \left\{D_1, \ldots, D_n\right\}$.

\subsection{Models}
We assume the conditional distribution of $\Ybf_{ig}$ to follow a Gaussian distribution. Therefore, we consider a linear mixed effects sub-model for each $\Ybf_{ig}$:
\begin{eqnarray}
\label{eq1}
\Ybf_{ig} &=& m_{ig}(t) + \epsilon_{ig}(t), \cr
&=& X_{ig}^{\top}(t)\beta_g + Z_{ig}^{\top}(t)b_{ig} + \epsilon_{ig}(t), \quad g=1,\ldots, G,   
\end{eqnarray}
where $m_{ig}(t)$ is the true mean trajectory of $\Ybf_{ig}$, $X_{ig}(t)$ and $Z_{ig}(t)$ denote the possible time-varying covariates, and $\beta_g$ and $b_{ig}$ represent the fixed effects and random effects for the $g^{th}$ longitudinal response. $\epsilon_{ig} (t)\sim N(0, \sigma_g^2)$ is the measurement error independent of $b_{ig}$, and $\epsilon_{ig} (t_1)$ is independent of $\epsilon_{ig} (t_2)$ for any $t_1 \neq t_2$. $\epsilon_{ig}$ is independent of $\epsilon_{ig'}$ for any $g \neq g'$. The joint vector of $q$-dimensional random effects $\bbf_i = (b_{i1}^{\top},\ldots,b_{iG}^{\top})^{\top}$ are assumed to follow a multivariate normal distribution
$MVN(0, \Sigma)$ with mean 0 and variance-covariance matrix
\begin{eqnarray*}
\Sigma =  \left( {\begin{array}{ccc}
   \Sigma_{11} & ... & \Sigma_{1G} \\
   \vdots & \ddots & \vdots  \\
   \Sigma_{G1} & ... & \Sigma_{GG}
  \end{array} } \right),
\end{eqnarray*}
where $\Sigma_{gg'}=cov(b_{ig},b_{ig'}) \forall g, g' \in [G]$.

Assume further that the competing risks time-to-event outcome follows the cause-specific Cox proportional hazards sub-model. All $G$ longitudinal responses are associated with the competing risk outcome through the random effects $b_{ig}, g=1,\ldots,G$:
\begin{eqnarray}
\label{eq2}
\lambda_{ik}(t) & = & \lambda_{0k}(t) \exp\left(W_i^{\top} \gamma_k + \sum_{g=1}^G \alpha_{kg}^{\top}b_{ig}\right),\quad k=1,\ldots, K,
\end{eqnarray}
where  $\lambda_{0k}(t \mid s)$ is a completely unspecified baseline hazard function, $W_i$ is a vector of baseline covariates, with its associated fixed effects $\gamma_k$. The two sub-models (\ref{eq1}) and (\ref{eq2}) are linked together by the latent association structure $\sum_{g=1}^G \alpha_{kg}^{\top}b_{ig}$, with $\alpha_{kg}$ a vector of the association parameter that quantifies the strength of association between the $g^{th}$ longitudinal response and the $k^{th}$ failure-type time-to-event outcome. Of note, we opt for this shared random effects parameterization as it uses the random effects as features extracted from the subject-specific mean and variance components to influence the survival outcome. It might lead to less interpretable association parameters, specifically when the spline functions are considered as random effects covariates \citep{rizopoulos2012joint,lawrence2015joint}. On the other hand, this time-independent association structure is useful for dynamic prediction of the survival outcome, and it opens a new path to facilitate efficient implementation of the cumulative baseline hazard function from a computational perspective \citep{lawrence2015joint, li2022efficient}. In Section \ref{sec:linearscan}, we will discuss efficient fitting of the shared random effects model for large data.

\subsection{Semiparametric maximum likelihood estimation via an EM algorithm}\label{Sect:EM}
Let $\Psi = \left\{\bm \beta, \sigma_1^2, \ldots,\sigma_G^2, \gamma_1, \ldots, \gamma_K, \alpha_{1}, \ldots, \alpha_{K}, \Sigma, \lambda_{01}(\cdot), ..., \lambda_{0K}(\cdot)\right\}$ denote all unknown parameters for the joint model (\ref{eq1}) and (\ref{eq2}), where $\bm \beta = (\beta_1^{\top}, ..., \beta_G^{\top})^{\top}$,  and $\alpha_k = (\alpha_{k1}^{\top}, ..., \alpha_{kG}^{\top})^{\top}$. Omitting the covariates for the sake of brevity, the observed-data likelihood is given by
\begin{eqnarray}
\label{likelihood}
    \prod_{i=1}^n f(\Ybf_i, T_i, D_i \mid \Psi)& = &\prod_{i=1}^n \int f(\Ybf_i \mid \bbf_i, \Psi)f(T_i, D_i \mid \bbf_i, \Psi)f(\bbf_i \mid \Psi)d\bbf_i\\ 
    &=& \prod_{i=1}^n \int \left\{\prod_{g=1}^G f(\Ybf_{ig} \mid b_{ig}, \Psi)\right\}f(T_i, D_i \mid \bbf_i, \Psi)f(\bbf_i \mid \Psi)d\bbf_i,
\end{eqnarray}
where 
\begin{eqnarray*}
    f(\Ybf_{ig} \mid \bbf_i, \Psi) &=& \prod_{j=1}^{n_{ig}} \frac{1}{\sqrt{2\pi\sigma_g^2}}\exp \left[-\frac{1}{2\sigma_g^2}\left\{ Y_{ig}(t_{ijg})-X_{ig}^{\top}(t_{ijg})\beta_g - Z_{ig}^{\top}(t_{ijg})b_{ig}\right\}^2\right], \\
    f(T_i, D_i \mid \bbf_i, \Psi) &=& \prod_{k=1}^K \left\{\lambda_{0k}(T_i) \exp\left(W_i^{\top} \gamma_k + \sum_{g=1}^G \alpha_{kg}^{\top} b_{ig}\right) \right\}^{I(D_i=k)} \cr
    &&\times \exp\left\{- \sum_{k=1}^K \lambda_{0k}(T_i)\exp\left(W_i^{\top} \gamma_k + \sum_{g=1}^G \alpha_{kg}^{\top} b_{ig}\right) \right\}, \\
    f(\bbf_i \mid \Psi) &=&  \frac{1}{\sqrt{(2\pi)^q\mid\Sigma\mid}}\exp(-\frac{1}{2}\bbf_i^{\top} \Sigma^{-1} \bbf_i),
\end{eqnarray*}
where the first equality follows from the assumption that $Y_i$ and $(T_i, D_i)$ are independent conditional on the covariates and the random effects. 

Because $\Psi$ contains $K$ unknown cause-specific baseline hazard functions and the likelihood function involves integrals,  directly maximizing the above observed-data likelihood is difficult. To tackle this issue, we derive an EM algorithm to compute the semi-parametric maximum likelihood estimate (SMLE) of $\Psi$ by regarding the latent random effects $\bbf_i$ as missing data \citep{dempster1977maximum, elashoff2008joint}.The complete-data likelihood based on $(\Ybf, \Tbf, \Dbf, \bbf)$ is given by
\begin{eqnarray*}
     L(\Psi; \Ybf, \Tbf, \Dbf, \bbf) &\propto& \prod_{i=1}^n \prod_{j=1}^{n_{ig}} \frac{1}{\sqrt{2\pi\sigma_g^2}}\exp \left[-\frac{1}{2\sigma_g^2}\left\{ Y_{ig}(t_{ijg})-X_{ig}^{\top}(t_{ijg})\beta_g - Z_{ig}^{\top}(t_{ijg})b_{ig}\right\}^2\right] \cr
     && \times \prod_{k=1}^K \left\{ \Delta \Lambda_{0k}(T_i) \exp\left(W_i^{\top} \gamma_k + \sum_{g=1}^G \alpha_{kg}^{\top} b_{ig}\right) \right\}^{I(D_i=k)} \cr
     &&\times \exp\left\{- \sum_{k=1}^K \Lambda_{0k}(T_i)\exp\left(W_i^{\top} \gamma_k + \sum_{g=1}^G \alpha_{kg}^{\top} b_{ig}\right) \right\}, \\
    && \times \frac{1}{\sqrt{(2\pi)^{q}\mid \Sigma\mid }}\exp\left(-\frac{1}{2}\bbf_i^{\top} \Sigma^{-1} \bbf_i\right),
\end{eqnarray*}
where $\Lambda_{0k}(.)$ is the cumulative baseline hazards function for type $k$ failure and
$\Delta \Lambda_{0k}(T_i)= \Lambda_{0k}(T_i) - \Lambda_{0k}(T_i-)$, and $q=\sum_{g=1}^G q_g$ the dimension of the random effects $\bbf_i$.

The EM algorithm iterates between an expectation step (E-step):
\begin{eqnarray}
\label{MstepFunction}
Q(\Psi;\Psi^{(m)}) &\equiv& \sum_{i=1}^n E_i^{(m)} l_i (\Ybf_i, T_i, D_i \mid \Psi) \cr
&=& \sum_{i=1}^n E_i^{(m)}\left[\left\{ \sum_{g=1}^G \log f(\Ybf_{ig} \mid \bbf_i, \Psi) \right\} + \log f(T_i, D_i \mid \bbf_i, \Psi) + \log f(\bbf_i \mid \Psi)\right],
\end{eqnarray}
 and a maximization step (M-step):
 \begin{equation} \label{Mstep}
    \Psi^{(m+1)} = \arg\max_{\Psi} Q(\Psi;\Psi^{(m)}),
\end{equation}
until the algorithm converges, where $\Psi^{(m)}$ is the estimate of $\Psi$ from the $m$-th iteration.
The E-step in (\ref{MstepFunction}) is taken with respect to the conditional expectation of the random effects $\bbf_i$ on the observed data $\left\{\Ybf_i, T_i, D_i\right\}$ at the current set of estimate of $f(\bbf_i \mid \Ybf_i, T_i, D_i, \Psi^{(m)})$. The M-step in (\ref{Mstep}) includes the expectations of some functional forms of $\bbf_i$ in the preceding E-step, say $h(\bbf_i)$, whose expectation is calculated with respect to the posterior distribution $f(\bbf_i \mid \Ybf_i, T_i, D_i, \Psi^{(m)})$.

\subsection{Standard error estimation}
\label{sec:computational-details-SE}
As discussed in \cite{elashoff2016joint} (Section 4.1, p.72), several approaches including profile-likelihood, observed information matrix, and bootstrap method have been proposed in the literature for estimating the standard errors of the parametric components of the SMLE. Here we adopt the profile-likelihood approach because it can be readily computed from the EM algorithm and performed well in our simulation studies.  Moreover, its computational burden can be further reduced to $O(n)$ using the customized linear scan algorithms introduced by \citet{li2022efficient}.  

Let $\Omega = (\bm{\beta}, \sigma_1^2, \ldots,\sigma_G^2, \gamma_1, \ldots, \gamma_K, \alpha_{1}, \ldots, \alpha_{K}, \Sigma)$ denote the parametric component of $\Psi$ and $\hat{\Omega}$ its SMLE. We propose to estimate the variance-covariance matrix of $\hat{\Omega}$ by inverting the empirical Fisher information obtained from the profile likelihood of $\Omega$ \citep{lin2004latent, zeng2005asymptotic, zeng2005simultaneous} as follows:
\begin{equation}
\label{SEestimation}
    \sum_{i=1}^n [\nabla_{\Omega}l^{(i)}(\hat{\Omega})][\nabla_{\Omega} l^{(i)}(\hat{\Omega})]^{\top},
\end{equation}
where $\nabla_{\Omega}l^{(i)}(\hat{\Omega})$ is the gradient vector of the conditional expectation of the complete data (profile) log-likelihood function evaluated at $\hat{\Omega}$ for subject $i$. Detailed formulas for obtaining the observed score vector via normal approximation for each parametric component are provided in Section \ref{Appen:computational-details-SE}, equations (\ref{Appen:SEbeta}) through (\ref{Appen:SEnu}), of the Supplementary Materials.

\subsection{Computational aspects}
\subsubsection{An approximate EM algorithm via normal approximation}
Gauss-Hermite quadrature \citep{press2007numerical,naylor1982applications} is one of the most widely used numerical integration techniques within the EM algorithm framework. However, it becomes computationally burdensome in the presence of high-dimensional random effects. To address this challenge in the context of jointly modeling multiple longitudinal biomarkers, recent work by \citet{bernhardt2015fast,murray2022fast,murray2023fast} proposed approximating the conditional distribution of random effects, given the observed data and current parameter estimates $\Psi^{(m)}$, with a (multivariate) normal distribution,
\begin{eqnarray}
\label{biappro}
    \bbf_i \mid \Ybf_i, T_i, D_i, \Psi^{(m)} \overset{\cdot}{\sim} N(\hat{\bbf}_i, \hat{\Sigma}_i).
\end{eqnarray}
Here $\hat{\bbf}_i$ is the posterior mode of $f(\Ybf_i, T_i, D_i, \bbf_i \mid \Psi^{(m)})$ at the current EM iteration, with the estimated covariance
\begin{eqnarray*}
    \hat{\Sigma}_i = \left\{ -\frac{\partial^2 \log f(\Ybf_i, T_i, D_i, \bbf_i \mid \Psi^{(m)})}{\partial \bbf_i \partial \bbf_i ^{\top}} \Big |_{\bbf_i = \hat{\bbf}_i} \right\}^{-1}.
\end{eqnarray*}
The normal approximation of the conditional distribution of $\bbf_i$ was previously demonstrated in \citet{rizopoulos2012fast} as $n_{ig} \rightarrow \infty$, and \citet{bernhardt2015fast,murray2022fast,murray2023fast} also demonstrated that the parameter estimation appears reasonable even for fewer longitudinal follow-up measures. Following the approximation (\ref{biappro}), any linear combination of $\bbf_i$ is also normal, thus their expectations in (\ref{MstepFunction}) can be calculated analytically in approximation (\ref{biappro}). As for the nonlinear terms in (\ref{MstepFunction}), \citet{bernhardt2015fast,murray2022fast,murray2023fast} used normal approximation in combination with Gauss Hermite quadrature method with three abscissas to evaluate the expectations in a one-dimensional integration, such as $E_i\left\{\exp(\alpha_k^{\top} \bbf_{i})\right\}$ when updating the baseline hazard $\lambda_{0k}(t)$ and fixed effects $\gamma_k$ in the survival sub-model. Compared to Gauss-Hermite quadrature, the hybrid of normal approximation and Gauss-Hermite quadrature substantially relaxes the computational burden since it reduces the the dimension of integration in the E-step of the algorithm to one, regardless of the number of random effects in the joint model. 


In our EM algorithm, the expectations of the nonlinear terms involved in (\ref{MstepFunction}) can be approximated via moment generating function (MGF). For instance, following the approximation (\ref{biappro}), evaluating $E_i\left[\exp\left\{ \alpha_k^{\top} \bbf_{i}\right\}\right]$ required in updating $\lambda_{0k}(.)$, $\gamma_k$, and $\alpha_k$ can be approximated by the MGF of $\bbf_{i}$, $M_{\bbf_{i}}(t) = \exp\left\{\hat{\bbf}_{i}^{\top} t + \frac{1}{2}t^{\top} \hat{\Sigma}_i t\right\}$ evaluated at $t = \alpha_k$. Following this idea, the two other expectations $E_i\left\{\bbf_{i}\exp(\alpha_k^{\top} \bbf_{i})\right\}$ and $E_i\left\{\bbf_{i} \bbf_{i}^{\top}\exp(\alpha_k^{\top} \bbf_{i})\right\}$ can be approximated by the first and second derivative of $M_{\bbf_{i}}(t)$ evaluated at $t = \alpha_k$. Details of derivations of these terms are provided in Section \ref{Appen:Mstepdetails} of Supplementary Materials. Since there are no other nonlinear terms involved in our EM algorithm relative to the necessary terms appearing in \citet{bernhardt2015fast,murray2022fast,murray2023fast}, Gauss-Hermite quadrature is no longer needed in evaluating all these three nonlinear terms, which can further relax the computational burden. In our simulation results, we demonstrated that our algorithms can still yield satisfactory estimation accuracy and coverage without requiring any numerical integration.

\subsubsection{Linear scan algorithms: scalable parameter and standard error estimation}
\label{sec:linearscan}
We note that our EM algorithm and standard error estimation method developed for the joint model (\ref{eq1})–(\ref{eq2}) requires many double summations when evaluating the quantities over risk set due to the non-parametric $\lambda_{0k}(.)$, with each necessitating
$O(n^2)$ evaluations of exponential functions when implemented naively, which are computationally expensive. This can lead to significant computational bottlenecks, particularly when the EM algorithm is slow to converge. Moreover, the formulas for standard error estimation outlined in Section \ref{Appen:computational-details-SE} of the Supplementary Materials also involve double summations for each subject, requiring  $O(n^3)$ evaluations of exponential functions overall. However, since the survival sub-model includes only time-independent covariates and shared random effects, we can reduce the computational complexity to $O(n)$ using linear scan algorithms similar to \citet{li2022efficient}. In Section \ref{Appen:Comp} of the Supplementary Materials, we provide details on these linear scan algorithms used for both the EM algorithm and standard error estimation in our joint model with time-independent covariates and shared random effects.

\section{Simulations}
\label{Sec:simulations}
\subsection{Parameter estimation}
We considered a simulation study with a five-variate joint model, where $Y_{ig}(t_{ijg})$ was generated by a linear mixed effects model:
\begin{eqnarray}
\label{sim:eq1}
    Y_{ig}(t_{ijg}) &=& \beta_{g0} + \beta_{g1}X_{i1} + \beta_{g2}X_{i2} + \beta_{g3}t_{ijg} + b_{ig0} + b_{ig1}t_{ijg} + \epsilon_{ig}(t_{ijg}), \cr
    && \epsilon_{ig}(t_{ijg}) \sim N(0, \sigma_g^2), \quad g=1,\ldots,5,
\end{eqnarray}
and the competing risks event data were generated from a cause-specific Cox's proportional hazards model:
\begin{eqnarray}
\label{sim:eq2}
    \lambda_{ik}(t) &=& \lambda_{0k}(t) \exp\left\{\gamma_{k1}X_{i1} + \gamma_{k2}X_{i2} + \sum_{g=1}^5 (\alpha_{kg1} b_{ig0} + \alpha_{kg2} b_{ig1})\right\}, \quad k=1,2.
\end{eqnarray}
Here, we set $X_{i1}$ as a Bernoulli draw ($p=0.5$), $X_{i2}$ a continuous variable drawn from $Uni(-5,5)$. We set the fixed effects $\beta_1 = (5, 1.5, 2, 1)^{\top}$, $\beta_2 = (10, 1, 2, 1)^{\top}$, $\beta_3 = (10, -2, 1, 0.5)^{\top}$, $\beta_4 = (7, -2, 2, 1)^{\top}$, $\beta_5 = (5, 1, 2, 2)^{\top}$. $t_{ijg}$ was a series of follow-up time points $(0, 0.7, 1.4,\ldots)$ with an increment of 0.7 across all subjects and biomarkers, which controlled the average length of longitudinal profile to be approximately 7 per subject. The longitudinal responses were simulated with their corresponding variances $\sigma_g^2 = 0.5$ induced by independent random errors $\epsilon_{ig}(t_{ijg}),g=1,\ldots,5$. The random effects are 10-dimensional, $\bbf_i = (b_{i10}, b_{i11}, b_{i20}, b_{i21},b_{i30}, b_{i31},b_{i40}, b_{i41},b_{i50}, b_{i51})^{\top} \sim N(0, \Sigma)$, with $\diag(\Sigma)=(1,1,1,1,1,1,1,1,1,1)^{\top}$ and all non-diagonal elements in $\Sigma$ are zero. The fixed effects in the survival sub-model (\ref{sim:eq2}) were $\gamma_1 = (1, 0.5)$, $\gamma_2 = (-0.5, 0.5)$, and the association parameters were $\alpha_{11} = \alpha_{21} = (0.5, 0.7)$, $\alpha_{12} = \alpha_{22} = (-0.5, 0.5)$, $\alpha_{13} = \alpha_{23} = (0.1, 0.5)$, $\alpha_{14} = \alpha_{24} = (-0.1, 0.4)$, $\alpha_{15} = \alpha_{25} = (0.2, 0.3)$. The baseline hazards $\lambda_{01}(t), \lambda_{02}(t)$ are set to constants 0.05 and 0.025, respectively. Independent, non-informative censoring times $C_i$ were drawn from $Uni(4,8)$. Let $T_i = \min(T_{i1}^*, T_{i2}^*, C_i)$ be the observed survival time (possibly right-censored), where $T_{i1}^*$ and $T_{i2}^*$ are independent conditional on the covariates $X_i$, $\bbf_i$, from models (\ref{sim:eq1}) and (\ref{sim:eq2}), respectively, $i=1,\ldots, n$. The longitudinal measurements for subject $i$ are assumed missing when $t_{ijg}>T_i$. The median censoring rate is 47.2\%, and the median event rates are 27.8\% and 25.0\% for type 1 and type 2 failures.

We evaluated the estimation performance of our estimation algorithms for the joint model (\ref{sim:eq1})-(\ref{sim:eq2}) Table \ref{sim:tab1} summarizes the simulation results, including bias, empirical standard deviations of the parameter estimates (SD), average estimated standard errors (SE), and coverage probabilities (CP, \%) of the 95\% confidence intervals, based on 300 Monte Carlo replicates with a sample size of 
$n=800$.  

It is observed from Table \ref{sim:tab1} that the proposed joint model (Model 1) performs well, exhibiting small biases in all parameter and standard error estimates, with CP close to the nominal 95\% level.

\begin{center}
\label{sim:tab1}
\begin{longtable}{llccccc}
\caption{Parameter estimates for the five-variate simulation scenario. Bias, empirical standard deviation (SD), 
median estimated standard error (SE), and coverage probabilities (CP) for each parameter were calculated from each model fit of all 300 Monte Carlo replicates ($n$ = 800).} \\
\toprule
Parameter & Truth & Bias & SD & SE & CP (\%) \\
\midrule
\endfirsthead
\multicolumn{6}{c}%
{{\bfseries \tablename\ \thetable\ -- continued}} \\
\toprule
Parameter & Truth & Bias & SD & SE & CP (\%) \\
\midrule
\endhead
\midrule \multicolumn{6}{r}{{Continued on next page}} \\
\endfoot
\bottomrule
\endlastfoot
Longitudinal &  &    & &  & \\
$\beta_{10}$ & 5 & -0.03 & 0.06 & 0.06 & 95.7 \\
$\beta_{11}$ & 1.5 & -0.01 & 0.07 & 0.09 & 98.7 \\
$\beta_{12}$ & 2 & -0.01 & 0.01 & 0.01 & 91.0 \\
$\beta_{13}$ & 1 & -0.02 & 0.03 & 0.05 & 97.3 \\
$\beta_{20}$ & 10 & 0.03 & 0.06 & 0.06 & 94.3 \\
$\beta_{21}$ & 1 & 0.01 & 0.08 & 0.09 & 96.7 \\
$\beta_{22}$ & 2 & 0.01 & 0.01 & 0.01 & 92.3 \\
$\beta_{23}$ & 1 & -0.01 & 0.02 & 0.05 & 99.7 \\
$\beta_{30}$ & 10 & -0.01 & 0.06 & 0.06 & 96.3 \\
$\beta_{31}$ & -2 & 0.01 & 0.08 & 0.09 & 96.3 \\
$\beta_{32}$ & 1 & $<$0.01 & 0.01 & 0.01 & 97.0 \\
$\beta_{33}$ & 0.5 & -0.01 & 0.02 & 0.05 & 99.7 \\
$\beta_{40}$ & 7 & $<$0.01 & 0.05 & 0.06 & 96.3 \\
$\beta_{41}$ & -2 & 0.01 & 0.08 & 0.09 & 96.7 \\
$\beta_{42}$ & 2 & $<$0.01 & 0.01 & 0.01 & 96.7 \\
$\beta_{43}$ & 1 & -0.01 & 0.02 & 0.05 & 99.7 \\
$\beta_{50}$ & 5 & -0.02 & 0.06 & 0.06 & 94.7 \\
$\beta_{51}$ & 1 & $<$0.01 & 0.08 & 0.09 & 96.7 \\
$\beta_{52}$ & 2 & $<$0.01 & 0.01 & 0.01 & 96.0 \\
$\beta_{53}$ & 2 & $<$0.01 & 0.02 & 0.05 & 100 \\
$\sigma^2_1$ &0.5 &$<$0.01 &0.01 &0.01 &95.0\\
$\sigma^2_2$ &0.5 &$<$0.01 &0.01 &0.01 &94.3\\
$\sigma^2_3$ &0.5 &$<$0.01 &0.01 &0.01 &95.0\\
$\sigma^2_4$ &0.5 &$<$0.01 &0.01 &0.01 &95.7\\
$\sigma^2_5$ &0.5 &$<$0.01 &0.01 &0.01 &93.0\\
$\Sigma_{11}$ &1 &$<$0.01 &0.06 &0.07 &97.0\\
$\Sigma_{22}$ & 1 & 0.03 & 0.07 & 0.07 & 95.7 \\
$\Sigma_{33}$ & 1 & 0.01 & 0.06 & 0.07 & 98.7 \\
$\Sigma_{44}$ & 1 & 0.01 & 0.07 & 0.07 & 97.7 \\
$\Sigma_{55}$ & 1 & $<$0.01 & 0.06 & 0.07 & 97.3 \\
$\Sigma_{66}$ & 1 & 0.01 & 0.07 & 0.07 & 94.7 \\
$\Sigma_{77}$ & 1 & -0.01 & 0.06 & 0.07 & 95.3 \\
$\Sigma_{88}$ & 1 & $<$0.01 & 0.06 & 0.07 & 96.0 \\
$\Sigma_{99}$ & 1 & $<$0.01 & 0.06 & 0.07 & 98.0 \\
$\Sigma_{10,10}$ & 1 & $<$0.01 & 0.06 & 0.07 & 97.0 \\
Survival &  &    & &  & \\
$\gamma_{11}$ & 1 & -0.02 & 0.16 & 0.18 & 98.0 \\
$\gamma_{12}$ & 0.5 & -0.04 & 0.03 & 0.04 & 90.0 \\
$\gamma_{21}$ & -0.5 & -0.03 & 0.15 & 0.18 & 97.0 \\
$\gamma_{22}$ & 0.5 & -0.03 & 0.04 & 0.04 & 92.0 \\
$\alpha_{111}$ & 0.5 & -0.05 & 0.09 & 0.10 & 95.3 \\
$\alpha_{112}$ & 0.7 & -0.06 & 0.09 & 0.12 & 96.7 \\
$\alpha_{121}$ & -0.5 & 0.06 & 0.08 & 0.10 & 95.3 \\
$\alpha_{122}$ & 0.5 & -0.05 & 0.09 & 0.12 & 97.3 \\
$\alpha_{131}$ & 0.1 & -0.01 & 0.08 & 0.10 & 98.3 \\
$\alpha_{132}$ & 0.5 & -0.06 & 0.09 & 0.12 & 96.7 \\
$\alpha_{141}$ & -0.1 & 0.01 & 0.08 & 0.10 & 98.7 \\
$\alpha_{142}$ & 0.4 & -0.05 & 0.09 & 0.12 & 97.7 \\
$\alpha_{151}$ & 0.2 & -0.02 & 0.09 & 0.10 & 96.7 \\
$\alpha_{152}$ & 0.3 & -0.04 & 0.09 & 0.11 & 97.7 \\
$\alpha_{211}$ & 0.5 & -0.04 & 0.09 & 0.11 & 95.3 \\
$\alpha_{212}$ & 0.7 & -0.05 & 0.10 & 0.13 & 96.0 \\
$\alpha_{221}$ & -0.5 & 0.06 & 0.09 & 0.11 & 95.3 \\
$\alpha_{222}$ & 0.5 & -0.05 & 0.10 & 0.12 & 97.0 \\
$\alpha_{231}$ & 0.1 & 0.00 & 0.09 & 0.10 & 97.3 \\
$\alpha_{232}$ & 0.5 & -0.04 & 0.10 & 0.12 & 99.0 \\
$\alpha_{241}$ & -0.1 & 0.02 & 0.09 & 0.10 & 97.7 \\
$\alpha_{242}$ & 0.4 & -0.04 & 0.09 & 0.12 & 99.3 \\
$\alpha_{251}$ & 0.2 & -0.02 & 0.08 & 0.10 & 97.3 \\
$\alpha_{252}$ & 0.3 & -0.02 & 0.10 & 0.12 & 98.0 \\
\end{longtable}
\end{center}

\subsection{Scalability}
\begin{figure}[htbp]
  \centering
  \includegraphics[width=1\textwidth]{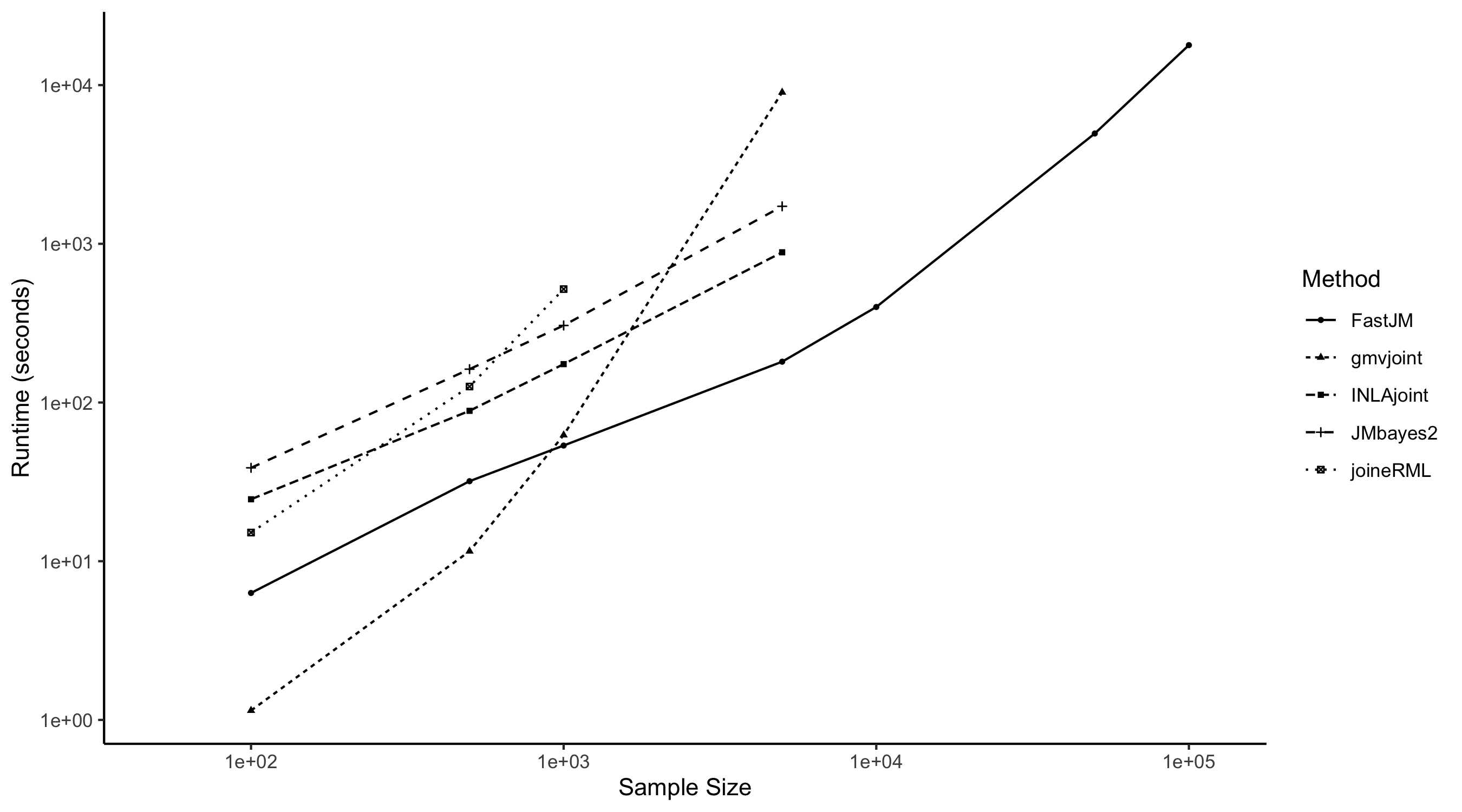}
  \caption{Average run time versus sample size of different joint model software packages for $G=2$ longitudinal biomarkers based on 5 Monte Carlo replicates.}
 \label{Fig:runtime}
\end{figure}

We presented a simulation study to illustrate the computational efficiency of our joint modeling package, \texttt{FastJM}, across various sample sizes. For comparison purposes, we included four existing R packages: \texttt{joineRML} \citep{hickey2018joinerml}, \texttt{gmvjoint} \citep{murray2022fast,murray2023fast}, \texttt{JMbayes2} \citep{rizopoulos2024optimizing}, and \texttt{INLAjoint} \citep{rustand2024fast}. Table~\ref{tab:jmpackages} summarizes the key features of these packages. We evaluated computational performance in fitting joint models with $G = 2$ longitudinal biomarkers and a single time-to-event outcome. Figure~\ref{Fig:runtime} displays the average run times under increasing sample sizes $n = {100, 500, 1000, 5000, 10000, 50000, 100000}$, based on 5 Monte Carlo replicates. All simulations were performed on a MacBook Pro with an M4 Pro chip and 24 GB unified memory, with parallel computation disabled to ensure fair comparisons across packages.

Among the methods, \texttt{gmvjoint} exhibited the fastest performance for small datasets (e.g., 1.14s for $n = 100$ and 11.56s for $n = 500$) but became the slowest at $n = 5000$, likely due to the non-linear computational complexity introduced by its nonparametric baseline hazard estimation. \texttt{joineRML} was faster than \texttt{JMbayes2} and \texttt{INLAjoint} with a sample size of 100 but its computational time grew non-linearly as the sample size increased and finally encountered memory overload when $n = 5000$. \texttt{INLAjoint} exhibited a smaller run time compared to \texttt{JMbayes2} because of its use of Empirical Bayes rather than fully Bayesian estimation, but it still spent 4 times more time finishing a run than \texttt{FastJM} with a sample size of 5000 (884.29s versus 180.81s). Although both \texttt{JMbayes2} and \texttt{INLAjoint} exhibited linear runtime growth, \texttt{JMbayes2} failed to complete a run at $n = 10000$ due to memory limitations, and \texttt{INLAjoint} also failed to produce parameter estimates at $n = 10000$ as it faced convergence failure across all replicates. Although \texttt{INLAjoint} does not require as much memory for model fitting as \texttt{JMbayes2}, its memory usage increases substantially with larger sample sizes, as noted by its developers \citep{rustand2024joint}. In contrast, \texttt{FastJM} can continue to finish a model fit with a reasonable runtime and tolerable memory consumption in a larger sample size setting ($n \geq 10000$), with nearly linear growth in computational time. This simulation demonstrated a consistently strong computational performance of \texttt{FastJM}, offering clear advantages over the other four packages in terms of both speed and memory efficiency.
\begin{table}[ht]
\centering
\resizebox{\textwidth}{!}{
\begin{tabular}{ccccc}
\hline
Packages & Baseline hazard & Competing risks & Latent association structure & Estimation algorithms \\
\hline
\texttt{FastJM} & Unspecified & Yes & Shared random effects & Frequentist \\
\texttt{INLAjoint} & Poisson & Yes & Shared random effects & Bayesian-INLA \\
 &  &  & Current value/slope & \\
\texttt{gmvjoint} & Unspecified & No & Current value of the latent process & Frequentist \\
\texttt{JMbayes2} & B-spline & Yes & Current value/slope/area & Bayesian-MCMC \\

\texttt{joineRML} & Unspecified & No & Current value of the latent process & MCEM \\
\hline
\end{tabular}
}\caption{Summary of the joint model software packages compared}
\label{tab:jmpackages}
\end{table}

\section{Example: Mayo Clinic Primary Biliary Cirrhosis Data}
\label{Sec:application}
We used the \textbf{pbc2} dataset as provided in the R package \texttt{JMbayes2}, including 312 randomized patients with PBC followed at the Mayo Clinic between 1974 and 1988 \citep{murtaugh1994primary}. We considered 5 longitudinal markers: log platelets per cubic ml/1000, log serum bilirubin in mg/dl, log SGOT (aspartate aminotransferase in U/ml), albumin in g/dl, prothrombine time in seconds. Of the 312 patients, 140 (44.9\%) were dead during the follow-up and 29 (9\%) received liver transplantation, in which we consider as a competing risks event. The maximum follow-up time is 14.3 years with a number of individual repeated measurements ranging between 1
and 16 with a median of 5. This dataset has been widely used to illustrate joint modeling approaches for multivariate longitudinal and survival data \citep{philipson2020faster,devaux2022individual,murray2022fast,murray2023fast}, and later account for the presence of competing risks \citep{rustand2024fast}.

\begin{figure}[htbp]
  \centering
  \includegraphics[width=0.9\textwidth]{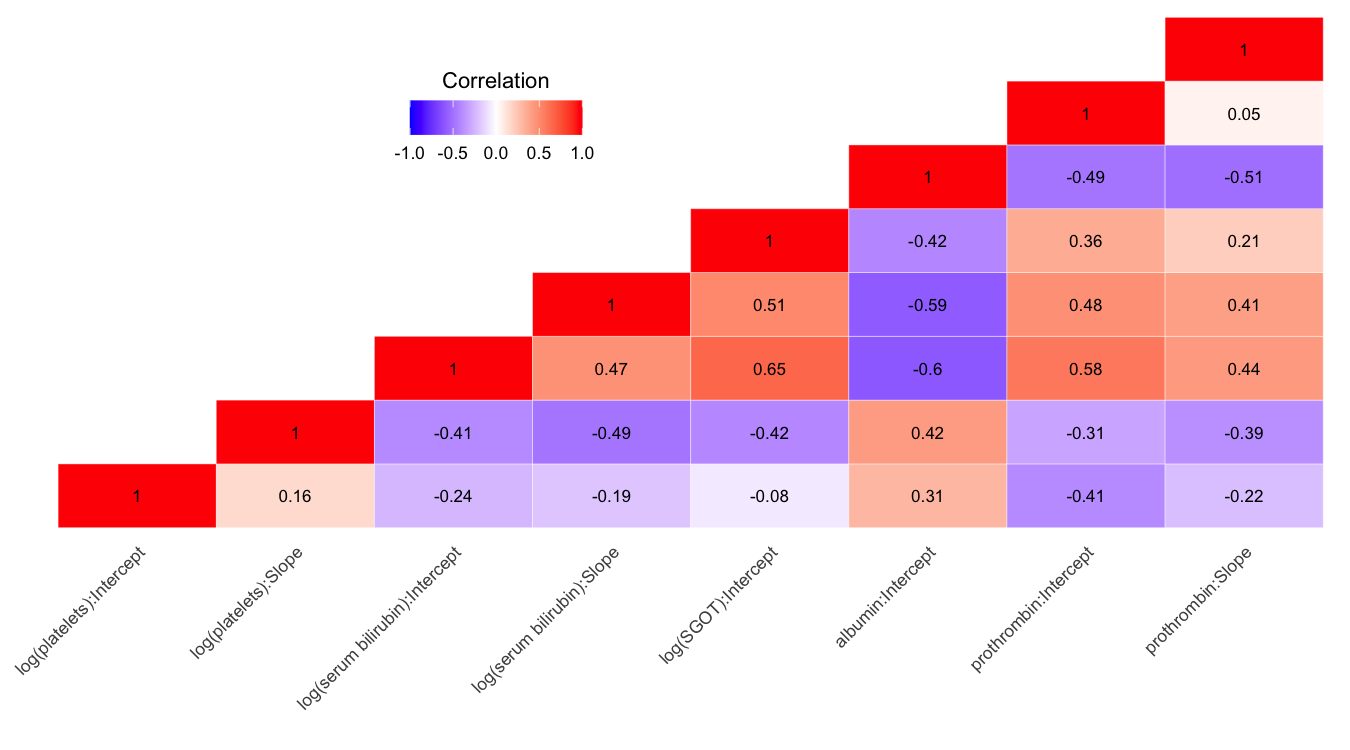}
  \caption{Matrix of correlations of the random effects between the 5 longitudinal markers calculated from the \texttt{FastJM} package.}
  \label{Fig:pbc2cor}
\end{figure}

As an illustrative example, we consider a multivariate joint model for these 5 biomarkers:
\begin{eqnarray*}
    \log Y_{i1}(t) &=& (\beta_{10} + b_{i10}) + (\beta_{11} + b_{i11}) t + \beta_{12}\text{drug}_i + \epsilon_{i1}(t),\cr
    \log Y_{i2}(t) &=& (\beta_{20} + b_{i20}) + (\beta_{21} + b_{i21}) t + \beta_{22}\text{drug}_i + \epsilon_{i2}(t),\cr
    \log Y_{i3}(t) &=& (\beta_{30} + b_{i30}) + \beta_{31} t + \beta_{32}\text{drug}_i + \epsilon_{i3}(t),\cr
    Y_{i4}(t) &=& (\beta_{40} + b_{i40}) + \beta_{41} t + \beta_{42}\text{drug}_i + \epsilon_{i4}(t),\cr
    Y_{i5}(t) &=& (\beta_{50} + b_{i50}) + (\beta_{51} + b_{i50})  t + \beta_{52}\text{drug}_i + \epsilon_{i5}(t),\cr
    \lambda_{ik}(t) &=& \lambda_{0k}(t) \exp\left(\gamma_k \text{drug}_i + \sum_{g=1}^5 \alpha_{kg}^{\top} b_{ig}\right), \quad k = 1, 2,
\end{eqnarray*}
where $b_{ig} = (b_{ig0}, b_{ig1})^{\top}$ for $g=1,2,5$, and $b_{ig} = b_{ig0}$ for $g=3,4$. The first, second, and fifth mixed effects sub-models included both the fixed and random intercept and slope, which corresponds to platelets (log), serum bilirubin (log), and prothrombine. The third and forth mixed effects sub-models included the fixed effects and random intercept only, associated with SGOT (log) and albumin. The survival sub-models are Cox proportional hazards models for the risk of death ($k=1$) and liver transplantation ($k=2$). The association between the longitudinal and competing risks outcome are modeled through the shared random effects $\sum_{g=1}^5 \alpha_{kg}^{\top} b_{ig}$. We acknowledge that various modeling strategies could have been employed; however, our goal here is to provide a rationale for the selected biomarkers based on their clinical relevance, rather than relying on a formal model selection procedure. The parameter estimates are summarized in Table \ref{tab:pbc2}, and the correlations of the random effects between the 5 biomarkers are calculated and plotted in Figure \ref{Fig:pbc2cor}. Of note, these results are for illustrative purposes only as some parameter estimates exhibited large values and standard errors, potentially due to limited sample size and large number of hyperparameters from the non-parametric baseline hazards.

\begin{center}
\begin{longtable}{llcc}
\caption{Parameter estimates for application to \textbf{pbc2} dataset. The total computation time for our estimation algorithms was 3.5 minutes. We also tried an existing R package \texttt{INLAjoint} \citep{rustand2024fast}, and its computation time was 17 minutes. The time-invariant survival fixed effects parameters $\gamma_k, k=1,2$ are not associated with a specific longitudinal outcome, as such they are reported separately. The variance-covariance matrix $\Sigma_{gg}$ is reported for each of the responses in the form $\Sigma_{gg,ef}$, where $g$ denotes the longitudinal biomarker and $e,f$ the random-effects indices. The results of the non-block diagonal variance-covariance matrices $\Sigma_{gg'}, g' \neq g,$ are omitted here.} \\
\toprule
Outcome & Parameter
& Estimate (SE) & 95\% CI \\
\midrule
\endfirsthead
\multicolumn{4}{c}%
{{\bfseries \tablename\ \thetable\ -- continued}} \\
\toprule
Outcome & Parameter
& Estimate (SE) & 95\% CI  \\
\midrule
\endhead
\midrule \multicolumn{4}{c}{{Continued on next page}} \\
\endfoot
\bottomrule
\endlastfoot
log(platelets) & $\beta_{10}$ &5.505 (0.047) & [5.413, 5.598] \\
 & $\beta_{11}$ &-0.063 (0.007) & [-0.077, -0.049] \\
  & $\beta_{12}$ &-0.078 (0.060) & [-0.196, 0.040] \\
& $\sigma_1^2$ &0.043 (0.001) & [0.042, 0.045] \\
& $\Sigma_{11, 00}$&0.139 (0.015) & [0.109, 0.169] \\
& $\Sigma_{11, 01}$&0.003 (0.002) & [-0.002, 0.008] \\
&$\Sigma_{11, 11}$ &0.003 (0.001) & [0.001, 0.004]\\ 
& $\alpha_{11}$ (death) & &\\
& random intercept &-0.093 (0.518) & [-1.108, 0.923]\\
& random slope &1.021 (4.202) & [-7.215, 9.256]\\
& $\alpha_{21}$ (transplantation) & &\\
& random intercept &0.124 (1.33) & [-2.484, 2.732]\\
& random slope &-6.718 (12.7) & [-31.61, 18.174]\\
& & & \\
log(serum bilirubin) & $\beta_{20}$ &0.538 (0.122) & [0.298, 0.777] \\
 & $\beta_{21}$ &0.182 (0.02) & [0.141, 0.222] \\
  & $\beta_{22}$ &-0.18 (0.153) & [-0.481, 0.12] \\
& $\sigma_2^2$ &0.122 (0.001) & [0.12, 0.124] \\
& $\Sigma_{22, 00}$&0.993 (0.169) & [0.662, 1.323] \\
& $\Sigma_{22, 01}$&0.077 (0.026) & [0.027, 0.128] \\
&$\Sigma_{22, 11}$ &0.03 (0.006) & [0.018, 0.042] \\
& $\alpha_{12}$ (death) & &\\
& random intercept &0.035 (0.348) & [-0.646, 0.717]\\
& random slope &0.303 (1.743) & [-3.113, 3.718]\\
& $\alpha_{22}$ (transplantation) & &\\
& random intercept &1.988 (1.032) & [-0.034, 4.01]\\
& random slope &8.219 (4.05) & [0.281, 16.158]\\
& & & \\
log(SGOT) & $\beta_{30}$ &4.798 (0.059) & [4.683, 4.913] \\
 & $\beta_{31}$ &-0.004 (0.002) & [-0.008, 0] \\
  & $\beta_{32}$ &-0.173 (0.075) & [-0.32, -0.027] \\
& $\sigma_3^2$ &0.094 (0.002) & [0.091, 0.097] \\
& $\Sigma_{33, 00}$&0.204 (0.032) & [0.142, 0.266]\\
& $\alpha_{13}$ (death) & &\\
& random intercept &0.052 (0.607) & [-1.139, 1.242]\\
& $\alpha_{23}$ (transplantation) & &\\
& random intercept &-1.56 (1.849) & [-5.184, 2.064]\\
& & & \\
albumin & $\beta_{40}$ &3.511 (0.056) & [3.401, 3.621] \\
 & $\beta_{41}$ &-0.077 (0.003) & [-0.082, -0.071] \\
  & $\beta_{42}$ &0.039 (0.065) & [-0.088, 0.166] \\
& $\sigma_4^2$ &0.122 (0.002) & [0.118, 0.126] \\
& $\Sigma_{44, 00}$&0.143 (0.024) & [0.095, 0.19]\\
& $\alpha_{14}$ (death) & &\\
& random intercept &-0.324 (0.727) & [-1.749, 1.102]\\
& $\alpha_{24}$ (transplantation) & &\\
& random intercept &-0.936 (2.068) & [-4.989, 3.117]\\
& & & \\
 prothrombin & $\beta_{50}$ &10.658 (0.138) & [10.388, 10.929] \\
 & $\beta_{51}$ &0.293 (0.058) & [0.178, 0.408] \\
  & $\beta_{52}$ &-0.174 (0.169) & [-0.506, 0.158] \\
& $\sigma_5^2$ &1.107 (0.019) & [1.07, 1.145] \\
& $\Sigma_{55, 00}$&0.68 (0.143) & [0.399, 0.96] \\
& $\Sigma_{55, 01}$&0.013 (0.037) & [-0.059, 0.085] \\
&$\Sigma_{55, 11}$ &0.102 (0.028) & [0.047, 0.157]\\
& $\alpha_{15}$ (death) & &\\
& random intercept &0.092 (0.467) & [-0.824, 1.008]\\
& random slope &-0.011 (1.439) & [-2.833, 2.81]\\
& $\alpha_{25}$ (transplantation) & &\\
& random intercept &-1.561 (1.174) & [-3.862, 0.74]\\
& random slope &-1.127 (4.108) & [-9.179, 6.924]\\
& & & \\
&$\gamma_1$ (death) &-2.177 (0.233) & [-2.633, -1.721] \\
&$\gamma_2$ (transplantation) &-0.412 (0.569) & [-1.528, 0.704]
\label{tab:pbc2}
\end{longtable}
\end{center}

\section{Discussion}
\label{Sec:discussion}
In this article, we present an efficient implementation of a semiparametric joint model for multivariate longitudinal biomarkers and competing risks time-to-event data, along with a publicly available R package, \texttt{FastJM}. Our estimation algorithms alleviate the computational burden associated with 1) multi-dimensional integration over random effects and 2) the non-linear growth in complexity in both parameter and standard error estimation associated with hyperparameters of nonparametric baseline hazards. This is achieved by incorporating a normal approximation approach \citep{murray2022fast, murray2023fast} and customized linear scan algorithms \citep{li2022efficient}, thus facilitating the application of multivariate joint models in large-scale datasets.

We demonstrate that our estimation algorithms outperform existing methods in terms of computational efficiency, yielding accurate estimates with reduced run-time and manageable memory usage. Compared to other software packages for multivariate joint modeling, \texttt{FastJM} offers several notable advantages. First, it imposes no restriction on the number of longitudinal biomarkers that can be included in the joint model. Second, it accommodates competing risks data, as demonstrated in our application study—a feature currently not supported by \texttt{rstanarm}, \texttt{joineRML}, and \texttt{gmvjoint}. Third, among the R packages that do support competing risks (e.g., \texttt{JMbayes2}, \texttt{INLAjoint}), \texttt{FastJM} is built upon a frequentist framework, which eliminates the need for prior specification, offers a flexible modeling structure capable of handling a large number of hyperparameters, and requires significantly less computation time and memory compared to Bayesian approaches. This makes \texttt{FastJM} especially well-suited for large-scale data settings.

We acknowledge several limitations of \texttt{FastJM} as currently implemented. First, the distributional assumption for longitudinal biomarkers is limited to Gaussian outcomes. However, incorporating non-Gaussian outcomes (e.g., binary, Poisson) into multivariate joint models is also important, particularly as such outcomes are frequently collected in biobank-scale databases and are often correlated with clinical endpoints. This capability is currently supported by other R packages such as \texttt{JMbayes2}, \texttt{INLAjoint}, and \texttt{gmvjoint}. Extending our methodology to accommodate non-Gaussian outcomes is an active area of future research and will be addressed in a forthcoming paper. Second, the customized linear scan algorithms used in our estimation procedure do not support time-dependent survival covariates or latent association structures. Of note, \texttt{FastJM} only supports shared random effects parameterizations of the form $\sum_{g=1}^G \alpha_{kg}^{\top}b_{ig}$. As discussed in Section~\ref{Sec:methods}, latent association structures may take various forms with distinct interpretations. However, the primary objective of this work is to enable fast fitting of joint models using a suitable association structure in the context of large-scale data.

\section*{Conflict of interest}
Not declared.

\section*{Software}
A user-friendly R package \texttt{FastJM} to fit a multivariate joint model developed in this paper is publicly available at The Comprehensive R Archive Network \url{https://CRAN.R-project.org/package=FastJM}.

\bibliographystyle{apalike}
\bibliography{reference}

\begin{thebibliography}{}

\bibitem[Bernhardt et~al., 2015]{bernhardt2015fast}
Bernhardt, P.~W., Zhang, D., and Wang, H.~J. (2015).
\newblock A fast em algorithm for fitting joint models of a binary response and
  multiple longitudinal covariates subject to detection limits.
\newblock {\em Computational statistics \& data analysis}, 85:37--53.

\bibitem[Crowther et~al., 2013]{crowther2013joint}
Crowther, M.~J., Abrams, K.~R., and Lambert, P.~C. (2013).
\newblock Joint modeling of longitudinal and survival data.
\newblock {\em The Stata Journal}, 13(1):165--184.

\bibitem[Dempster et~al., 1977]{dempster1977maximum}
Dempster, A.~P., Laird, N.~M., and Rubin, D.~B. (1977).
\newblock Maximum likelihood from incomplete data via the em algorithm.
\newblock {\em Journal of the Royal Statistical Society: Series B
  (Methodological)}, 39(1):1--22.

\bibitem[Devaux et~al., 2022]{devaux2022individual}
Devaux, A., Genuer, R., Peres, K., and Proust-Lima, C. (2022).
\newblock Individual dynamic prediction of clinical endpoint from large
  dimensional longitudinal biomarker history: a landmark approach.
\newblock {\em BMC Medical Research Methodology}, 22(1):188.

\bibitem[Elashoff et~al., 2016]{elashoff2016joint}
Elashoff, R., Li, N., et~al. (2016).
\newblock {\em Joint modeling of longitudinal and time-to-event data}.
\newblock CRC Press.

\bibitem[Elashoff et~al., 2008]{elashoff2008joint}
Elashoff, R.~M., Li, G., and Li, N. (2008).
\newblock A joint model for longitudinal measurements and survival data in the
  presence of multiple failure types.
\newblock {\em Biometrics}, 64(3):762--771.

\bibitem[Elashoff et~al., 2017]{ElashoffLiLi17JMBook}
Elashoff, R.~M., Li, G., and Li, N. (2017).
\newblock {\em Joint Modeling of Longitudinal and Time-to-Event Data}, volume
  151 of {\em Monographs on Statistics and Applied Probability}.
\newblock CRC Press, Boca Raton, FL.

\bibitem[Goodrich et~al., 2024]{rstanarm2024}
Goodrich, B., Gabry, J., Ali, I., and Brilleman, S. (2024).
\newblock {\em rstanarm: Bayesian Applied Regression Modeling via Stan}.
\newblock R package version 2.32.1.

\bibitem[Henderson et~al., 2000]{henderson2000joint}
Henderson, R., Diggle, P., and Dobson, A. (2000).
\newblock Joint modelling of longitudinal measurements and event time data.
\newblock {\em Biostatistics}, 1(4):465--480.

\bibitem[Hickey et~al., 2016]{hickey2016joint}
Hickey, G.~L., Philipson, P., Jorgensen, A., and Kolamunnage-Dona, R. (2016).
\newblock Joint modelling of time-to-event and multivariate longitudinal
  outcomes: recent developments and issues.
\newblock {\em BMC medical research methodology}, 16:1--15.

\bibitem[Hickey et~al., 2018]{hickey2018joinerml}
Hickey, G.~L., Philipson, P., Jorgensen, A., and Kolamunnage-Dona, R. (2018).
\newblock joinerml: a joint model and software package for time-to-event and
  multivariate longitudinal outcomes.
\newblock {\em BMC medical research methodology}, 18:1--14.

\bibitem[Lawrence~Gould et~al., 2015]{lawrence2015joint}
Lawrence~Gould, A., Boye, M.~E., Crowther, M.~J., Ibrahim, J.~G., Quartey, G.,
  Micallef, S., and Bois, F.~Y. (2015).
\newblock Joint modeling of survival and longitudinal non-survival data:
  current methods and issues. report of the dia bayesian joint modeling working
  group.
\newblock {\em Statistics in medicine}, 34(14):2181--2195.

\bibitem[Li et~al., 2022]{li2022efficient}
Li, S., Li, N., Wang, H., Zhou, J., Zhou, H., and Li, G. (2022).
\newblock Efficient algorithms and implementation of a semiparametric joint
  model for longitudinal and competing risk data: with applications to massive
  biobank data.
\newblock {\em Computational and mathematical methods in medicine},
  2022(1):1362913.

\bibitem[Lin et~al., 2004]{lin2004latent}
Lin, H., McCulloch, C.~E., and Rosenheck, R.~A. (2004).
\newblock Latent pattern mixture models for informative intermittent missing
  data in longitudinal studies.
\newblock {\em Biometrics}, 60(2):295--305.

\bibitem[Murray and Philipson, 2022]{murray2022fast}
Murray, J. and Philipson, P. (2022).
\newblock A fast approximate em algorithm for joint models of survival and
  multivariate longitudinal data.
\newblock {\em Computational Statistics \& Data Analysis}, 170:107438.

\bibitem[Murray and Philipson, 2023]{murray2023fast}
Murray, J. and Philipson, P. (2023).
\newblock Fast estimation for generalised multivariate joint models using an
  approximate em algorithm.
\newblock {\em Computational statistics \& data analysis}, 187:107819.

\bibitem[Murtaugh et~al., 1994]{murtaugh1994primary}
Murtaugh, P.~A., Dickson, R.~E., Van~Dam, G.~M., Malinchoc, M., Grambsch,
  P.~M., Langworthy, A.~L., and Gips, C.~H. (1994).
\newblock Primary biliary cirrhosis: prediction of short--term survival based
  on repeated patient visits.
\newblock {\em Hepatology}, 20(1):126--134.

\bibitem[Naylor and Smith, 1982]{naylor1982applications}
Naylor, J.~C. and Smith, A.~F. (1982).
\newblock Applications of a method for the efficient computation of posterior
  distributions.
\newblock {\em Journal of the Royal Statistical Society Series C: Applied
  Statistics}, 31(3):214--225.

\bibitem[Philipson et~al., 2020]{philipson2020faster}
Philipson, P., Hickey, G.~L., Crowther, M.~J., and Kolamunnage-Dona, R. (2020).
\newblock Faster monte carlo estimation of joint models for time-to-event and
  multivariate longitudinal data.
\newblock {\em Computational Statistics \& Data Analysis}, 151:107010.

\bibitem[Press et~al., 2007]{press2007numerical}
Press, W.~H., Teukolsky, S.~A., Vetterling, W.~T., and Flannery, B.~P. (2007).
\newblock {\em Numerical recipes 3rd edition: The art of scientific computing}.
\newblock Cambridge university press.

\bibitem[Putter et~al., 2007]{putter2007tutorial}
Putter, H., Fiocco, M., and Geskus, R.~B. (2007).
\newblock Tutorial in biostatistics: competing risks and multi-state models.
\newblock {\em Statistics in medicine}, 26(11):2389--2430.

\bibitem[Rizopoulos, 2010]{Rizopoulos10JM}
Rizopoulos, D. (2010).
\newblock {JM}: An {R} package for the joint modelling of longitudinal and
  time-to-event data.
\newblock {\em Journal of Statistical Software}, 35(9):1--33.

\bibitem[Rizopoulos, 2012a]{rizopoulos2012fast}
Rizopoulos, D. (2012a).
\newblock Fast fitting of joint models for longitudinal and event time data
  using a pseudo-adaptive gaussian quadrature rule.
\newblock {\em Computational Statistics \& Data Analysis}, 56(3):491--501.

\bibitem[Rizopoulos, 2012b]{rizopoulos2012joint}
Rizopoulos, D. (2012b).
\newblock {\em Joint models for longitudinal and time-to-event data: With
  applications in R}.
\newblock CRC press.

\bibitem[Rizopoulos and Taylor, 2024]{rizopoulos2024optimizing}
Rizopoulos, D. and Taylor, J.~M. (2024).
\newblock Optimizing dynamic predictions from joint models using super
  learning.
\newblock {\em Statistics in Medicine}, 43(7):1315--1328.

\bibitem[Rustand et~al., 2024a]{rustand2024joint}
Rustand, D., van Niekerk, J., Krainski, E.~T., and Rue, H. (2024a).
\newblock Joint modeling of multivariate longitudinal and survival outcomes
  with the r package inlajoint.
\newblock {\em arXiv preprint arXiv:2402.08335}.

\bibitem[Rustand et~al., 2024b]{rustand2024fast}
Rustand, D., Van~Niekerk, J., Krainski, E.~T., Rue, H., and Proust-Lima, C.
  (2024b).
\newblock Fast and flexible inference for joint models of multivariate
  longitudinal and survival data using integrated nested laplace
  approximations.
\newblock {\em Biostatistics}, 25(2):429--448.

\bibitem[Sattar and Sinha, 2019]{sattar2019joint}
Sattar, A. and Sinha, S.~K. (2019).
\newblock Joint modeling of longitudinal and survival data with a covariate
  subject to a limit of detection.
\newblock {\em Statistical methods in medical research}, 28(2):486--502.

\bibitem[Song et~al., 2002]{SongDavidianTsiatis02SemiparamJM}
Song, X., Davidian, M., and Tsiatis, A.~A. (2002).
\newblock A semiparametric likelihood approach to joint modeling of
  longitudinal and time-to-event data.
\newblock {\em Biometrics}, 58(4):742--753.

\bibitem[Tsiatis and Davidian, 2004]{tsiatis2004joint}
Tsiatis, A. and Davidian, M. (2004).
\newblock Joint modeling of longitudinal and time-to-event data: an overview.
\newblock {\em Statistica Sinica}, 14(3):809--834.

\bibitem[Wang, 2006]{Wang06CorrectedScore}
Wang, C.~Y. (2006).
\newblock Corrected score estimator for joint modeling of longitudinal and
  failure time data.
\newblock {\em Statistica Sinica}, 16(1):235--253.

\bibitem[Wulfsohn and Tsiatis, 1997]{wulfsohn1997joint}
Wulfsohn, M.~S. and Tsiatis, A.~A. (1997).
\newblock A joint model for survival and longitudinal data measured with error.
\newblock {\em Biometrics}, pages 330--339.

\bibitem[Zeng and Cai, 2005]{zeng2005simultaneous}
Zeng, D. and Cai, J. (2005).
\newblock Simultaneous modelling of survival and longitudinal data with an
  application to repeated quality of life measures.
\newblock {\em Lifetime Data Analysis}, 11:151--174.

\bibitem[Zeng et~al., 2005]{zeng2005asymptotic}
Zeng, D., Cai, J., et~al. (2005).
\newblock Asymptotic results for maximum likelihood estimators in joint
  analysis of repeated measurements and survival time.
\newblock {\em The Annals of Statistics}, 33(5):2132--2163.

\end{thebibliography}

\newpage
\appendix
\setcounter{page}{1}
\numberwithin{equation}{section}
\numberwithin{figure}{section}
\numberwithin{table}{section}

\makeatletter   
 \renewcommand{\@seccntformat}[1]{Section~{\csname the#1\endcsname}.\hspace*{1em}}
\makeatother

\setcounter{table}{0}
\renewcommand{\thetable}{S\arabic{section}.\arabic{table}}
\renewcommand{\thefigure}{S\arabic{section}.\arabic{figure}}

\makeatletter
\renewcommand \thesection{S\@arabic\c@section}
\makeatother

\title{Supplementary Materials to ``Efficient Implementation of a Semiparametric Joint Model for Multivariate Gaussian Longitudinal Responses and Competing Risks Data: With Applications to UKB-Biobank Data''}

\section{M step details}
\label{Appen:Mstepdetails}
The observed data likelihood for subject $i$ is defined in (\ref{likelihood}). Since we seek to obtain $\hat{\Psi}$ via SMLE, we present the complete data log-likelihood $l_i (\Ybf_i, T_i, D_i \mid \Psi)$ required by the E-step in (\ref{MstepFunction}), i.e.
{\small
\begin{eqnarray}
\label{complikelihood}
 l_i (\Ybf_i, T_i, D_i \mid \Psi) &=& \sum_{i=1}^n \left[\left\{ \sum_{g=1}^G \log f(\Ybf_{ig} \mid \bbf_i, \Psi) \right\} + \log f(T_i, D_i \mid \bbf_i, \Psi) + \log f(\bbf_i \mid \Psi)\right].
\end{eqnarray}
}
Note that each of these terms in (\ref{complikelihood}) involves a disjoint set of parameters; thus, to maximize (\ref{complikelihood}), we can maximize the expectation of each term separately. In the following, we describe the normal approximation algorithm used to update the parameters from iteration $(m)$ to iteration $(m +1)$.
 
\subsection{Update for \texorpdfstring{$\bm{\beta}$}{beta}}
We first set the log-likelihood $l_i (\Ybf_i, T_i, D_i \mid \Psi)$ with respect to $\beta$,
\begin{eqnarray*}
    l_i(\bm \beta) \propto -\frac{1}{2} (\Ybf_i - \Xbf_i^{\top}\bm \beta - \Zbf_i^{\top} \bbf_i)^{\top} V_i^{-1} (\Ybf_i - \Xbf_i^{\top}\beta - \Zbf_i^{\top} \bbf_i),
\end{eqnarray*}
where $\Xbf_i = \bigoplus_{g=1}^G X_{ig}$, $\Zbf_i = \bigoplus_{g=1}^G Z_{ig}$, and $V_i = \bigoplus_{g=1}^G \sigma_g^2 I_{n_{ig}}$, with $\bigoplus$ denoting the direct matrix sum and $I_x$ denoting an $x \times x$ identity matrix.

Now We need to compute the expectation of the above log-likelihood with respect to the posterior distribution of $\mathbf{b}_i$.

\[
E_i[l_i(\bm \beta)] = E_i \left[ -\frac{1}{2} (\Ybf_i - \Xbf_i^{\top} \bm \beta - \Zbf_i^{\top} \bbf_i)^{\top} \mathbf{V}_i^{-1} (\Ybf_i - \Xbf_i^{\top} \bm \beta - \Zbf_i^{\top} \bbf_i) \right]
\]
The random variable is \( \bbf_i \), and the fixed parameter is \( \bm \beta \). Using the approximation (\ref{biappro}), we have the conditional distribution of \( \bbf_i \) given the observed data is:
\[
\bbf_i \mid \Ybf_i, T_i, D_i, \Psi^{(m)} \sim N(\hat{\bbf}_i, \hat{\Sigma}_i),
\]
where \( \hat{\bbf}_i \) is the posterior mean, and \( \hat{\Sigma}_i \) is the posterior covariance of \( \bbf_i \). Therefore,

\[\Ybf_i - \Xbf_i^{\top} \bm \beta - \Zbf_i^{\top} \bbf_i \mid \Ybf_i, T_i, D_i, \Psi^{(m)} \sim N(\Ybf_i - \Xbf_i^{\top} \bm \beta -\Zbf_i^{\top}  \hat{\bbf}_i, \Zbf_i^{\top}\hat{\Sigma}_i\Zbf_i),
\]
Then we have 
\begin{eqnarray*}
E_i[l_i(\bm \beta)] &=& -\frac{1}{2} \left[ 
\text{Tr}\{\mathbf{V}_i^{-1}\Zbf_i^{\top}\hat{\Sigma}_i\Zbf_i\}+(\Ybf_i - \Xbf_i^{\top} \bm \beta -\Zbf_i^{\top}  \hat{\bbf}_i)^{\top}\mathbf{V}_i^{-1}(\Ybf_i - \Xbf_i^{\top} \bm \beta -\Zbf_i^{\top}  \hat{\bbf}_i)\right]
\end{eqnarray*}
To find the estimator for $\bm{\beta}$, we take the derivative of the log-likelihood with respect to $\bm{\beta}$:
{\small
\begin{eqnarray*}
   \frac{\partial E_i[l_i(\bm \beta)]}{\partial \bm \beta} &=&-\frac{1}{2}\frac{\partial}{\partial \bm \beta}(\Ybf_i - \Xbf_i^{\top} \bm \beta - \Zbf_i^{\top} \hat{\bbf}_i)^{\top} \mathbf{V}_i^{-1} (\Ybf_i - \Xbf_i^{\top} \bm \beta - \Zbf_i^{\top} \hat{\bbf}_i)\\
 &=&  -\frac{1}{2}\frac{\partial}{\partial \bm \beta}\left\{(\Ybf_i - \Zbf_i^{\top} \hat{\bbf}_i)^{\top} \mathbf{V}_i^{-1} (\Ybf_i - \Zbf_i^{\top} \hat{\bbf}_i)-2 (\bm{\beta}^{\top}\Xbf_i)\mathbf{V}_i^{-1}(\Ybf_i - \Zbf_i^{\top} \hat{\bbf}_i)+ (\bm{\beta}^{\top} \Xbf_i)\mathbf{V}_i^{-1} (\Xbf_i^{\top} \bm{\beta})\right\} \\
&=&\frac{\partial}{\partial \bm{\beta}} \left[ -\frac{1}{2} (\bm{\beta}^{\top}\mathbf{X}_i) \mathbf{V}_i^{-1} ( \mathbf{X}_i^{\top} \bm{\beta}) - \frac{1}{2} \cdot (-2) (\bm{\beta}^{\top}\mathbf{X}_i)\mathbf{V}_i^{-1}(\Ybf_i - \Zbf_i^{\top} \hat{\bbf}_i) \right] \\
&=& \mathbf{X}_i  \mathbf{V}_i^{-1} (\mathbf{Y}_i- \mathbf{Z}_i^{\top} \hat{\mathbf{b}}_i-\mathbf{X}_i^{\top} \bm{\beta})
\end{eqnarray*}
}
Setting the derivative to zero:
\begin{eqnarray*}
\mathbf{X}_i \mathbf{V}_i^{-1} (\mathbf{Y}_i -\mathbf{Z}_i^{\top} \hat{\mathbf{b}}_i) &=& \mathbf{X}_i \mathbf{V}_i^{-1}\mathbf{X}_i^{\top}\bm{\beta}
\end{eqnarray*}



To derive the estimator for \(\bm{\beta}\), we sum this equation over all subjects \(i\) from 1 to \(n\). Thus:

\[
\sum_{i=1}^n \mathbf{X}_i \mathbf{V}_i^{-1} (\mathbf{Y}_i - \mathbf{Z}_i^{\top} \hat{\mathbf{b}}_i) = \sum_{i=1}^n \mathbf{X}_i \mathbf{V}_i^{-1} \mathbf{X}_i^{\top} \bm{\beta}
\]

Solving for $\bm{\beta}$ gives:

\[
\hat{\bm{\beta}} = \left(\sum_{i=1}^n \mathbf{X}_i \mathbf{V}_i^{-1} \mathbf{X}_i^{\top}\right)^{-1} \left\{\sum_{i=1}^n \mathbf{X}_i \mathbf{V}_i^{-1} \left(\mathbf{Y}_i - \mathbf{Z}_i^{\top} \hat{\mathbf{b}}_i\right)\right\}
\]

\subsection{\texorpdfstring{Update for $\sigma_g^2$}{Update for sigma\_g²}}
We set the log-likelihood w.r.t. $\sigma_g^2$,
\begin{eqnarray*}
    l_i(\sigma_g^2) \propto -\frac{n_{ig}}{2} \log \sigma_g^2 - \sum_{j=1}^{n_{ig}}\frac{1}{2\sigma_g^2} \left\{Y_{ig}(t_{ijg}) - X_{ig}^{\top}(t_{ijg}) \beta_g - Z_{ig}^{\top}(t_{ijg}) b_{ig}\right\}^2.
\end{eqnarray*}
Define $r_{ig}(t_{ijg}) = Y_{ig}(t_{ijg}) - X_{ig}^{\top}(t_{ijg}) \beta_g$. Then the expected value 
{\small
\begin{eqnarray*}
    E_i [l_i(\sigma_g^2)] &=& -\frac{n_{ig}}{2} \log \sigma_g^2 - \sum_{j=1}^{n_{ig}}\frac{1}{2\sigma_g^2} E_i \left\{r_{ig}(t_{ijg}) - Z_{ig}^{\top}(t_{ijg}) b_{ig}\right\}^2, \\
    &=& -\frac{n_{ig}}{2} \log \sigma_g^2 - \sum_{j=1}^{n_{ig}}\frac{1}{2\sigma_g^2} \left[r_{ig}^2(t_{ijg}) -  2 r_{ig}(t_{ijg}) Z_{ig}^{\top}(t_{ijg}) E_i(b_{ig}) + \text{Tr}\left\{Z_{ig}(t_{ijg})Z_{ig}^{\top}(t_{ijg}) E_i(b_{ig} b_{ig}^{\top})\right\}\right]
\end{eqnarray*}
}
The first derivative w.r.t. $\sigma_g^2$ is
\begin{eqnarray*}
\frac{\partial E_i [l_i(\sigma_g^2)]}{\partial \sigma_g^2} &=&
 -\frac{n_{ig}}{2\sigma_g^2} + \frac{1}{2\sigma_g^4} \sum_{j=1}^{n_{ig}} \left[r_{ig}^2(t_{ijg}) - 2 r_{ig}(t_{ijg}) Z_{ig}^{\top}(t_{ijg}) E_i(b_{ig}) +  \text{Tr}\left\{Z_{ig}(t_{ijg})Z_{ig}^{\top}(t_{ijg}) E_i(b_{ig} b_{ig}^{\top})\right\}\right].
\end{eqnarray*}
Equating the above equation to be 0 and solve for $\sigma_g^2$ for all $i=1,\ldots,n$, we obtain the close-form update
\begin{eqnarray*}
    \hat{\sigma}_g^2 =\frac{\sum_{i=1}^n \sum_{j=1}^{n_{ig}} \left[r_{ig}^2(t_{ijg}) -  2 r_{ig}(t_{ijg}) Z_{ig}^{\top}(t_{ijg}) E_i(b_{ig}) +  \text{Tr}\left\{Z_{ig}(t_{ijg})Z_{ig}^{\top}(t_{ijg}) E_i(b_{ig} b_{ig}^{\top})\right\}\right]}{\sum_{i=1}^n n_{ig}}. 
\end{eqnarray*}
Using the approximation (\ref{biappro}), we have $E_i(b_{ig}) = \hat{b}_{ig}$ and $Var(\bbf_i) = \hat{\Sigma}_{ig}$, where $\hat{b}_{ig}$ is the posterior mode w.r.t. $b_{ig}$, i.e., the random effects for the $g^{th}$ longitudinal response, and $\hat{\Sigma}_{ig}$ is the $(g,g)^{th}$ block matrix of $\hat{\Sigma}_i$.  This results in
\begin{eqnarray*}
    E_i(b_{ig} b_{ig}^{\top}) &=& Var(b_{ig}) + E_i(b_{ig}) E_i(b_{ig})^{\top} \\
    &=& \hat{\Sigma}_{ig} + \hat{b}_{ig} \hat{b}_{ig}^{\top}.
\end{eqnarray*}
Therefore, we have 
\begin{eqnarray*}
    \hat{\sigma}_g^2 =\frac{\sum_{i=1}^n \sum_{j=1}^{n_{ig}} \left[r_{ig}^2(t_{ijg}) -  2 r_{ig}(t_{ijg}) Z_{ig}^{\top}(t_{ijg}) \hat{b}_{ig} +  \text{Tr}\left\{Z_{ig}(t_{ijg})Z_{ig}^{\top}(t_{ijg}) (\hat{\Sigma}_{ig} + \hat{b}_{ig} \hat{b}_{ig}^{\top})\right\}\right]}{\sum_{i=1}^n n_{ig}}. 
\end{eqnarray*}

\subsection{\texorpdfstring{Update for $\Sigma$}{Update for Sigma}}
We have
\begin{eqnarray*}
    E_i [\log f(\bbf_i \mid \Sigma)] &=& E_i \left[-\frac{q}{2} \log(2\pi) - \frac{1}{2}\log \mid\Sigma\mid - \frac{1}{2}\bbf_i^{\top} \Sigma^{-1} \bbf_i\right] \\
    &=& -\frac{q}{2} \log(2\pi) + \frac{1}{2} \log\mid\Sigma^{-1}\mid - \frac{1}{2}\text{Tr}\left\{\Sigma^{-1} E_i(\bbf_i \bbf_i^{\top})\right\}.
\end{eqnarray*}
The first derivative of $E_i [\log f(\bbf_i \mid \Sigma)]$ w.r.t. $\Sigma$ is
\begin{eqnarray*}
    \frac{\partial E_i [\log f(\bbf_i \mid \Sigma)]}{\partial \Sigma} = -\frac{1}{2}\Sigma + \frac{1}{2} E_i(\bbf_i \bbf_i^{\top}).
\end{eqnarray*}
Equating the above equation to be 0, we obtain the close-form update
\begin{eqnarray*}
    \hat{\Sigma} = E_i(\bbf_i \bbf_i^{\top}).
\end{eqnarray*}
Using the approximation (\ref{biappro}), we have $E_i(\bbf_i) = \hat{\bbf}_i$ and $Var(\bbf_i) = \hat{\Sigma}_i$. This results in
\begin{eqnarray*}
    E_i(\bbf_i \bbf_i^{\top}) &=& Var(\bbf_i) + E_i(\bbf_i) E_i(\bbf_i)^{\top} \\
    &=& \hat{\Sigma}_i + \hat{\bbf}_i \hat{\bbf}_i^{\top}.
\end{eqnarray*}
Finally, we solve for $\Sigma$ for all $i=1,\ldots,n$
\begin{eqnarray*}
    \hat{\Sigma} = \frac{\sum_{i=1}^n (\hat{\Sigma}_i + \hat{\bbf}_i \hat{\bbf}_i^{\top})}{n}.
\end{eqnarray*}

\subsection{\texorpdfstring{Update for $\Lambda_{0k}(.)$}{Update for Lambda\_0k(.)}}
We have
\begin{eqnarray*}
  \log  f(T_i, D_i \mid \bbf_i, \Psi) &=& \sum_{k=1}^K \left[I(D_i=k)\left\{\log \Delta \Lambda_{0k}(T_i) + W_i^{\top} \gamma_k + \sum_{g=1}^G \alpha_{kg}^{\top} b_{ig}\right\} \right. \cr
  && \left. -  \Lambda_{0k}(T_i)\exp\left(W_i^{\top} \gamma_k + \sum_{g=1}^G \alpha_{kg}^{\top} b_{ig}\right)\right]  \cr
  &=& \sum_{k=1}^K\left[I(D_i=k)\left\{\log \Delta \Lambda_{0k}(T_i) + W_i^{\top} \gamma_k +  \alpha_k^{\top} \bbf_{i}\right\}\right.  \cr
  && \left.-  \Lambda_{0k}(T_i)\exp(W_i^{\top} \gamma_k +  \alpha_k^{\top} \bbf_{i})\right].
\end{eqnarray*}

We now set the log likelihood w.r.t. $\Delta \Lambda_{0k}(.)$
\begin{eqnarray*}
    l_i(\Delta \Lambda_{0k}) &\propto& I(D_i = k) \log \Delta \Lambda_{0k}(T_i) -  \Lambda_{0k}(T_i)\exp(W_i^{\top} \gamma_k +  \alpha_k^{\top} \bbf_{i}) \\
    E_i\left[l_i(\Delta \Lambda_{0k})\right] &=& I(D_i = k) \log \Delta \Lambda_{0k}(T_i) -  \Lambda_{0k}(T_i) E_i\left[\exp(W_i^{\top} \gamma_k +  \alpha_k^{\top} \bbf_{i})\right].
\end{eqnarray*}
Using the approximation (\ref{biappro}), we have 
\begin{eqnarray}
\label{EWb}
   W_i^{\top} \gamma_k +  \alpha_k^{\top} \bbf_{i} \sim N\left(W_i^{\top} \gamma_k + \alpha_k^{\top}\hat{\bbf}_{i}, \alpha_k^{\top} \hat{\Sigma}_i \alpha_k  \right) = N\left(\mu_{ik}, \tau_{ik}^2\right).
\end{eqnarray}

Based upon equation (\ref{EWb}), calculating $E_i\left[\exp\left\{ \alpha_k^{\top} \bbf_{i}\right\}\right]$ is equivalent to the moment generating function (MGF) $M_{\bbf_{i}}(t) = \exp\left\{\hat{\bbf}_{i}^{\top} t + \frac{1}{2}t^{\top} \hat{\Sigma}_i t\right\}$ evaluated at $t = \alpha_k$. Therefore, we can approximate the expectation above as
\begin{eqnarray*}
    E_i\left[l_i(\Delta \Lambda_{0k})\right] &\approx& \Tilde{E}_i\left[l_i(\Delta \Lambda_{0k})\right] \\
    &=& I(D_i = k) \log \Delta \Lambda_{0k}(T_i) -  \Lambda_{0k}(T_i) \exp\left(W_i^{\top} \gamma_k + \alpha_k^{\top}\hat{\bbf}_{i} + \frac{1}{2}\alpha_k^{\top} \hat{\Sigma}_i \alpha_k\right).
\end{eqnarray*}
Take the first derivative w.r.t. $\Delta \Lambda_{0k}(.)$, we obtain the closed-form update of $ \Lambda_{0k}(.)$
\begin{eqnarray}
\label{Mstep:lambda}
    \hat{\Lambda}_{0k}(t) = \sum_{l: s< t_{kl} \leq t}  \frac{d_{kl}}{\sum_{r \in R(t_{kl})} \exp\left(W_r^{\top} \gamma_k + \alpha_k^{\top} \hat{\bbf}_{r} + \frac{1}{2}\alpha_k^{\top} \hat{\Sigma}_r \alpha_k\right)},
\end{eqnarray}
where $R(t_{kl})$ is the risk set at the uncensored $k$-th failure time $t_{kl}$, and $d_{kl}$ is the number of type $k$ failures at $t_{kl}$, for $k=1, \ldots, K.$

\subsection{\texorpdfstring{Update for $\gamma$ and $\alpha$}{Update for gamma and alpha}}
Using the survival log-likelihood, we can calculate the expectation on the survival fixed effects $\phi_k = (\gamma_k^{\top}, \alpha_k^{\top})^{\top}$:
\begin{eqnarray*}
    l_i(\phi_k) &\propto&
    I(D_i = k) (W_i^{\top} \gamma_k +  \alpha_k^{\top} \bbf_{i}) -  \Lambda_{0k}(T_i) \exp(W_i^{\top} \gamma_k +  \alpha_k^{\top} \bbf_{i}), \cr
    E_i\left\{l_i(\phi_k)\right\} &=&  I(D_i = k) \left\{W_i^{\top} \gamma_k + E_i (\alpha_k^{\top} \bbf_{i})\right\} -  \Lambda_{0k}(T_i) \exp\left(W_i^{\top} \gamma_k\right) E_i \left\{\exp(\alpha_k^{\top} \bbf_{i})\right\},
\end{eqnarray*}
where $\Lambda_{0k}(.)$ will be substituted with $\hat{\Lambda}_{0k}(.)$ from (\ref{Mstep:lambda}).
 We note that there is no closed-form solution for $\phi_k$ and it will be updated via an one-step Newton-Raphson algorithm from the current estimate:
\begin{eqnarray*}
    \hat{\phi_k} &=& \phi_k + I_{\phi_k}^{-1}S_{\phi_k},\quad k=1,\ldots,K,
\end{eqnarray*}
where 
\begin{eqnarray}
\label{Mstepscorefunc}
    S_{\phi_k} &=& (S_{\gamma_k}^{T}, S_{\alpha_k}^{T})^{T}, \\
    \label{Mstepinfo}
    I_{\phi_k} &=& 
    \begin{pmatrix}
I_{\gamma_k} & I_{\gamma_k \alpha_k} \\
I_{\gamma_k \alpha_k}^{T} & I_{\alpha_k}
\end{pmatrix}.
\end{eqnarray}
The explicit formulas of all the quantities in (\ref{Mstepscorefunc}) - (\ref{Mstepinfo}) are given by
\begin{eqnarray*}
    S_{\gamma_k} &=& \sum_{i=1}^n I(D_i = k) W_i -  \Lambda_{0k}(T_i) \exp\left(W_i^{\top} \gamma_k\right)E_i\left\{\exp(\alpha_k^{\top} \bbf_{i})\right\}W_i, \cr
     S_{\alpha_k} &=& \sum_{i=1}^n I(D_i = k) E_i (\bbf_{i}) -  \Lambda_{0k}(T_i) \exp\left(W_i^{\top} \gamma_k\right)E_i\left\{\bbf_{i}\exp( \alpha_k^{\top} \bbf_{i})\right\},\cr
     I_{\gamma_k} &=& \sum_{i=1}^n \Lambda_{0k}(T_i) \exp\left(W_i^{\top} \gamma_k\right)E_i\left\{\exp(\alpha_k^{\top} \bbf_{i})\right\}W_iW_i^{\top},\cr
     I_{\alpha_k} &=& \sum_{i=1}^n   \Lambda_{0k}(T_i) \exp\left(W_i^{\top} \gamma_k\right) E_i\left\{\bbf_{i} \bbf_{i}^{\top}\exp(\alpha_k^{\top} \bbf_{i})\right\}, \cr
     I_{\gamma_k \alpha_k} &=& \sum_{i=1}^n \Lambda_{0k}(T_i) \exp\left(W_i^{\top} \gamma_k\right) W_iE_i\left\{\bbf_{i}\exp( \alpha_k^{\top} \bbf_{i})\right\}^{\top}.
\end{eqnarray*}
Using the approximation (\ref{EWb}), approximating $E_i\left\{\exp(\alpha_k^{\top} \bbf_{i})\right\}$ will be the same as in updating $\Lambda_{0k}(.)$. The two expected values $E_i\left\{\bbf_{i}\exp(\alpha_k^{\top} \bbf_{i})\right\}$ and $E_i\left\{\bbf_{i} \bbf_{i}^{\top}\exp(\alpha_k^{\top} \bbf_{i})\right\}$ can be approximated by the first and second derivative of $M_{\bbf_{i}}(t)$ evaluated at $t = \alpha_k$, respectively:
\begin{eqnarray*}
   E_i\left\{\bbf_{i}\exp( \alpha_k^{\top} \bbf_{i})\right\} &\approx& \exp\left(\alpha_k^{\top}\hat{\bbf}_{i} + \frac{1}{2}\alpha_k^{\top} \hat{\Sigma}_i \alpha_k\right)\left(\hat{\Sigma}_i \alpha_k + \hat{\bbf}_{i}\right),
\end{eqnarray*}
\begin{eqnarray*}
    E_i\left[\bbf_{i} \bbf_{i}^{\top}\exp\left\{ \alpha_k^{\top} \bbf_{i}\right\}\right] &\approx& \exp\left(\alpha_k^{\top}\hat{\bbf}_{i} + \frac{1}{2}\alpha_k^{\top} \hat{\Sigma}_i \alpha_k\right) \left\{\left(\hat{\Sigma}_i \alpha_k + \hat{\bbf}_{i}\right)\left(\hat{\Sigma}_i \alpha_k + \hat{\bbf}_{i}\right)^{\top} + \hat{\Sigma}_i\right\}.
\end{eqnarray*}

\section{Score function formulas for standard error estimation}
\label{Appen:computational-details-SE}
Note that $l^{(i)}(\Omega)$ is the profile likelihood obtained by profiling out the baseline hazard $\lambda_0(.)$ from the observed data likelihood. Calculating its gradient, however, is difficult because there is no explicit expression for the profile likelihood. Here we approximate $\nabla_{\Omega}l^{(i)}(\hat{\Omega})$ by the derivative of the profile expected complete-data log-likelihood, obtained in the last E-step when the EM algorithm has converged. The parametric components of the observed score function $\nabla_{\Omega}l^{(i)}(\hat{\Omega})$ in equation (\ref{SEestimation}) can be calculated via normal approximation:

{\footnotesize
\begin{eqnarray}
\label{Appen:SEbeta}
\nabla_{\beta_g}l^{(i)} (\hat{\Omega})\!\!\! &=& \!\!\! \frac{1}{\sigma_g^2}\sum_{j=1}^{n_{ig}} E_i\left\{Y_{ig}(t_{ijg})-X_{ig}^{\top}(t_{ijg})\beta_g - Z_{ig}^{\top}(t_{ijg}) b_{ig}\right\} X_{ig}(t_{ijg})\Bigg |_{\Omega = \hat{\Omega}}, \cr
&\approx& \frac{1}{\sigma_g^2} \sum_{j=1}^{n_{ig}} \left\{Y_{ig}(t_{ijg})-X_{ig}^{\top}(t_{ijg})\beta_g\right\}X_{ig}(t_{ijg}) - X_{ig}(t_{ijg}) Z_{ig}^{\top}(t_{ijg}) \hat{b}_{ig} \Bigg |_{\Omega = \hat{\Omega}} \\
\label{Appen:SEsigma2}
\nabla_{\sigma_g^2}l^{(i)} (\hat{\Omega})\!\!\! &=&\!\!\! \left[ \frac{1}{2\sigma_g^4} \sum_{j=1}^{n_{ig}} E_i\left\{Y_{ig}(t_{ijg})-X_{ig}^{\top}(t_{ijg})\beta_g - Z_{ig}^{\top}(t_{ijg}) b_{ig}\right\}^2 \! - \! \frac{n_{ig}}{2\sigma_g^2} \right]\Bigg |_{\Omega = \hat{\Omega}}, \cr
&\approx& \sum_{j=1}^{n_{ig}}\frac{\left[r_{ig}^2(t_{ijg}) -  2 r_{ig}(t_{ijg}) Z_{ig}^{\top}(t_{ijg}) \hat{b}_{ig} +  \text{Tr}\left\{Z_{ig}(t_{ijg})Z_{ig}^{\top}(t_{ijg}) (\hat{\Sigma}_{ig} + \hat{b}_{ig} \hat{b}_{ig}^{\top})\right\}\right]}{2\sigma_g^4} - \frac{n_{ig}}{2\sigma_g^2}, \\
\label{Appen:SESigma}
\nabla_{\Sigma}l^{(i)} (\hat{\Omega}) \!\!\!  &=& \!\!\! \frac{1}{2}\left[2\Sigma^{-1}E_i(\bbf_i \bbf_i^{\top}) \Sigma^{-1}\!\! -\!\!  \left\{\Sigma^{-1}E_i(\bbf_i \bbf_i^{\top}) \Sigma^{-1} \circ I\right\} \!\! -\!\! 2\Sigma^{-1}\! + \!\Sigma^{-1} \circ I\right]\Bigg |_{\Omega = \hat{\Omega}}, \cr
&\approx& \frac{1}{2}\left[2\Sigma^{-1} (\hat{\Sigma}_i + \hat{\bbf}_i \hat{\bbf}_i^{\top}) \Sigma^{-1}\!\! -\!\!  \left\{\Sigma^{-1} (\hat{\Sigma}_i + \hat{\bbf}_i \hat{\bbf}_i^{\top}) \Sigma^{-1} \circ I\right\} \!\! -\!\! 2\Sigma^{-1}\! + \!\Sigma^{-1} \circ I\right], \\
\label{Appen:SEgamma}
\nabla_{\gamma_k}l^{(i)}(\hat{\Omega})\!\!\! &=& \!\!\! I(D_i = k) \left[ W_i - \frac{\sum_{r \in R(T_i)} \exp\left(W_r^{\top} \gamma_k\right)  E_r\left\{\exp(\alpha_k^{\top}\bbf_{r})\right\}W_r}{\sum_{r \in R(T_i)} \exp\left(W_r^{\top} \gamma_k\right)  E_r\left\{\exp(\alpha_k^{\top}\bbf_{r})\right\}} \right] \cr
&& + \left( \sum_{j: t_{kl} \leq T_i} \frac{d_{kj}\sum_{r \in R(t_{kl})} \exp\left(W_r^{\top} \gamma_k\right)  E_r\left\{\exp(\alpha_k^{\top}\bbf_{r})\right\}W_r}{\left[\sum_{r \in R(t_{kl})} \exp\left(W_r^{\top} \gamma_k\right)  E_r\left\{\exp(\alpha_k^{\top}\bbf_{r})\right\}\right]^2} \right. \cr
&& \left. -\sum_{j: t_{kl} \leq T_i}\frac{d_{kj}}{\sum_{r \in R(t_{kl})} \exp\left(W_r^{\top} \gamma_k\right)  E_r\left\{\exp(\alpha_k^{\top}\bbf_{r})\right\}}W_i \right) \cr
&& \times \exp\left(W_i^{\top} \gamma_k\right)E_i\left\{\exp(\alpha_k^{\top}\bbf_{i})\right\}\Bigg |_{\Omega = \hat{\Omega}}, \cr
 &\approx& \!\!\! I(D_i = k) \left\{ W_i - \frac{\sum_{r \in R(T_i)} \exp\left(W_r^{\top} \gamma_k + \alpha_k^{\top} \hat{\bbf}_{r} + \frac{1}{2}\alpha_k^{\top} \hat{\Sigma}_r \alpha_k\right) W_r}{\sum_{r \in R(T_i)} \exp\left(W_r^{\top} \gamma_k + \alpha_k^{\top} \hat{\bbf}_{r} + \frac{1}{2}\alpha_k^{\top} \hat{\Sigma}_r \alpha_k\right)} \right\} \cr
&& + \left[ \sum_{j: t_{kl} \leq T_i} \frac{d_{kj}\sum_{r \in R(t_{kl})} \exp\left(W_r^{\top} \gamma_k + \alpha_k^{\top} \hat{\bbf}_{r} + \frac{1}{2}\alpha_k^{\top} \hat{\Sigma}_r \alpha_k\right)W_r}{\left\{\sum_{r \in R(t_{kl})} \exp\left(W_r^{\top} \gamma_k + \alpha_k^{\top} \hat{\bbf}_{r} + \frac{1}{2}\alpha_k^{\top} \hat{\Sigma}_r \alpha_k\right)\right\}^2} \right. \cr
&& \left. -\sum_{j: t_{kl} \leq T_i}\frac{d_{kj}}{\sum_{r \in R(t_{kl})} \exp\left(W_r^{\top} \gamma_k + \alpha_k^{\top} \hat{\bbf}_{r} + \frac{1}{2}\alpha_k^{\top} \hat{\Sigma}_r \alpha_k\right)}W_i \right] \cr
&& \times \exp\left(W_i^{\top} \gamma_k + \alpha_k^{\top} \bbf_{i} + \frac{1}{2}\alpha_k^{\top} \hat{\Sigma}_i \alpha_k\right)\Bigg |_{\Omega = \hat{\Omega}}, \\
\label{Appen:SEnu}
\nabla_{\alpha_k}l^{(i)}(\hat{\Omega}) &=& I(D_i = k)\left[ E_i\left(\bbf_{i}\right) - \frac{\sum_{r \in R(T_i)} \exp\left(W_r^{\top} \gamma_k\right) E_r\left\{\bbf_{r}\exp\left(\alpha_k^{\top}\bbf_{r}\right)\right\}}{\sum_{r \in R(T_i)} \exp\left(W_r^{\top} \gamma_k\right) E_r\left\{\exp(\alpha_k^{\top}\bbf_{r})\right\}}\right] \cr
&& + \left( \sum_{j: t_{kl} \leq T_i} \frac{d_{kj}\sum_{r \in R(t_{kl})} \exp\left(W_r^{\top} \gamma_k\right) E_r\left\{\bbf_{r}\exp\left(\alpha_k^{\top}\bbf_{r}\right)\right\}}{\left[\sum_{r \in R(t_{kl})} \exp\left(W_r^{\top} \gamma_k\right) E_r\left\{\exp(\alpha_k^{\top}\bbf_{r})\right\}\right]^2} E_i\left\{\exp(\alpha_k^{\top}\bbf_{i})\right\} \right. \cr
&& \left. - \sum_{j: t_{kl} \leq T_i}\frac{d_{kj}}{\sum_{r \in R(t_{kl})} \exp\left(W_r^{\top} \gamma_k\right) E_r\left\{\exp(\alpha_k^{\top}\bbf_{r})\right\}} E_i\left\{\bbf_{i}\exp\left(\alpha_k^{\top}\bbf_{i}\right)\right\} \right) \cr
&& \times \exp\left(W_i^{\top} \gamma_k\right)\Bigg |_{\Omega = \hat{\Omega}}, \cr
&\approx& I(D_i = k)\left\{\hat{\bbf}_{i} - \frac{\sum_{r \in R(T_i)} \exp\left(W_r^{\top} \gamma_k + \alpha_k^{\top} \hat{\bbf}_{r} + \frac{1}{2}\alpha_k^{\top} \hat{\Sigma}_r \alpha_k\right)\left(\hat{\Sigma}_r \alpha_k + \hat{\bbf}_{r}\right)}{\sum_{r \in R(T_i)} \exp\left(W_r^{\top} \gamma_k + \alpha_k^{\top} \hat{\bbf}_{r} + \frac{1}{2}\alpha_k^{\top} \hat{\Sigma}_r \alpha_k\right)}\right\} \cr
&& + \left( \sum_{j: t_{kl} \leq T_i} \frac{d_{kj}\sum_{r \in R(t_{kl})} \exp\left(W_r^{\top} \gamma_k + \alpha_k^{\top} \hat{\bbf}_{r} + \frac{1}{2}\alpha_k^{\top} \hat{\Sigma}_r \alpha_k\right)\left(\hat{\Sigma}_r \alpha_k + \hat{\bbf}_{r}\right)}{\left[\sum_{r \in R(t_{kl})} \exp\left(W_r^{\top} \gamma_k + \alpha_k^{\top} \hat{\bbf}_{r} + \frac{1}{2}\alpha_k^{\top} \hat{\Sigma}_r \alpha_k\right)\right]^2} \right. \cr
&& \times \left. \exp\left(\alpha_k^{\top} \hat{\bbf}_{i} + \frac{1}{2}\alpha_k^{\top} \hat{\Sigma}_i \alpha_k\right) \right. \cr
&& \left. - \sum_{j: t_{kl} \leq T_i}\frac{d_{kj}}{\sum_{r \in R(t_{kl})} \exp\left(W_r^{\top} \gamma_k + \alpha_k^{\top} \hat{\bbf}_{r} + \frac{1}{2}\alpha_k^{\top} \hat{\Sigma}_r \alpha_k\right)} \right.\cr
&& \times \left. \exp\left\{\alpha_k^{\top} \bbf_{i} + \frac{1}{2}\alpha_k^{\top} \hat{\Sigma}_i \alpha_k\right\}\left\{\hat{\Sigma}_i \alpha_k + \hat{\bbf}_{i}\right\} \right)  \exp\left(W_i^{\top} \gamma_k\right)\Bigg |_{\Omega = \hat{\Omega}}. 
\end{eqnarray}
}

\section{Linear scan algorithms for efficient implementation for the shared random effects model}
\label{Appen:Comp}
Here, we derive linear scan algorithms to reduce the computational burden using similar ideas to \citet{li2022efficient}. Below we discuss some linear scan algorithms for implementation of the EM steps and standard error estimation.  
\subsection{Linear scan for the E-step}
The E-step involves evaluating the expectations of $h(\bbf_i)$ at each EM iteration, which requires calculating $f(T_i, D_i\mid \bbf_i, \Psi^{(m)})$ across all subjects. Note that $f(T_i, D_i\mid \bbf_i, \Psi^{(m)})$ can be rewritten as 
\begin{eqnarray*}
f(T_i, D_i\mid \bbf_i, \Psi) & = & \prod_{k=1}^K \left\{\Delta\Lambda_{0k}(T_i) \exp\left(W_i^{\top} \gamma_k + \alpha_k^{\top} \bbf_i\right)\right\}^{I(D_i=k)} \cr
     &&\times  \exp\left\{-\sum_{k=1}^K \int_0^{T_i} \exp\left(W_i^{\top} \gamma_k + \alpha_k^{\top} \bbf_i\right)d\Lambda_{0k}(t)\right\} \cr
& = & \prod_{k=1}^K \left\{\Delta\Lambda_{0k}(T_i) \exp(W_i^{\top} \gamma_k + \alpha_k^{\top} \bbf_i)\right\}^{I(D_i=k)}  \cr
&& \quad \times \exp \left\{-\sum_{k=1}^K \Lambda_{0k}(T_i) \exp(W_i^{\top} \gamma_k + \alpha_k^{\top} \bbf_i) \right\}.
\end{eqnarray*}
For each subject $i$, calculating $\Lambda_{0k}(T_i)$ would involve $O(n)$ operations if a global search is performed to find an interval of two adjacent uncensored event times that contains $T_i$. Consequently, calculating all $\Lambda_{0k}(T_i)$'s will require $O(n^2)$ operations. Similar to \citet{li2022efficient}, by taking advantage of the fact that $\Lambda_{0k}(t)$ is a right-continuous and non-decreasing step function, we define the following a linear scan map 
{\small
\begin{eqnarray}
\label{eqn:Lscan}
\{\Lambda_{0k}(t_{k1}), \Lambda_{0k}(t_{k2}), \ldots \Lambda_{0k}(t_{kq_k}) \} \mapsto \{\Lambda_{0k}(T_{(1)}), \Lambda_{0k}(T_{(2)}), \ldots, \Lambda_{0k}(T_{(n)})\},
\end{eqnarray}
}
where $t_{k1} > \cdots > t_{kq_k}$ are scanned forward from the largest to the smallest, and for each $t_{kl}$, only a subset of the ranked observation times $T_{(i)}$ are scanned forward to calculate  $\Lambda_{0k}(T_{(i)})$ as follows
\begin{equation*}
  \Lambda_{0k}(T_{(i)}) =
    \begin{cases}
     \Lambda_{0k}(t_{k1}) , & \text{if $T_{(i)} \geq t_{k1}$,}\\
      \Lambda_{0k}(t_{k(l+1)}), & \text{if $T_{(i)} \in [t_{k(l+1)}, t_{kl}), $ for some $l\in\{1,\ldots, q_k-1 \}$},\\
      0, & \text{$T_{(i)}<t_{kq_k}$}.
    \end{cases}       
\end{equation*}
Consequently, the entire algorithm for calculating all $\Lambda_{0k}(T_i)$'s costs only $O(n)$ operations since the scanned $T_{(i)}$'s for different $t_{kl}$'s do not overlap.
  
\subsection{Linear risk set scan for the M-step}
Multiple quantities in (\ref{Mstep:lambda})-(\ref{Mstepinfo}) including the cumulative baseline hazard functions involve aggregating information over the risk set $R(t_{kl}) = \{r: T_r \geq t_{kl}\}$ at each uncensored event time $t_{kl}$, which are further aggregated across all $t_{kl}$'s. All subjects are scanned to determine the risk set $R(t_{kl})$ for all uncensored event times will require $O(n^2)$ operations. Specifically, to update $\Lambda_{0k}(t_{kl})$, $\gamma_k$ and $\alpha_k$, one needs to compute $\sum_{r \in R(t_{kl})} a_r$, where $a_r$ is any time-independent quantity defined in equations (\ref{Mstep:lambda})-(\ref{Mstepinfo}). The risk set $R(t_{k(l+1)})$ can be decomposed into two disjoint sets: 
\begin{eqnarray}
\label{Appen:MstepScan}
\sum_{r \in R(t_{k(l+1)})} a_{r} = \sum_{r \in R(t_{kl})} a_{r} +\sum_{\{r: T_{(r)} \in [t_{k(l+1)}, t_{kl})\}}  a_{r},
\end{eqnarray}
where the distinct uncensored event times $t_{k1}> \cdots>t_{kq_k}$ are arranged in a decreasing order. 
it is easy to see that calculating $\sum_{r \in R(t_{kl})} a_{r}$, $j=1, \ldots q_k$, takes $O(n)$ operations when 
$T_{(r)}$'s are scanned backward in time, by following the recursive formula (\ref{Appen:MstepScan}) where the subjects in $R(t_{kl})$ do not need to be scanned to calculate the second term.

\subsection{Linear risk set scan for standard error estimation}
Standard error estimation formula in (\ref{SEestimation}) relies on the observed score vectors from the profile likelihood where the baseline hazards are profiled out. It is seen from equations (\ref{Appen:SEgamma})-(\ref{Appen:SEnu}) that obtaining the observed score vectors $\nabla_{\gamma_k}l^{(i)}(\hat{\Omega})$ and $\nabla_{\alpha_k}l^{(i)}(\hat{\Omega})$ involve aggregating information either over $\{ r\in R(T_i) \}$ or over both $\{r \in R(t_{kl})\}$ and $\{ j: t_{kl} \le T_i\}$,  which can takes either $O(n)$ or $O(n^2)$ operations, respectively, if not optimized. As a result, the empirical Fisher information matrix can take $O(n^3)$ operations as it requires summing up the information across all subjects. Specifically, to calculate the gradient $\nabla_{\gamma_k}l^{(i)}(\hat{\Omega})$ and $\nabla_{\alpha_k}l^{(i)}(\hat{\Omega})$, one needs to compute
\begin{eqnarray*}
    B(T_i) = \sum_{l: t_{kl} \le T_i}b_{kl}, \quad\mbox{for $i=1,..., n$,}
\end{eqnarray*}
where $B(.)$ is a right-continuous non-decreasing step function and $b_{kl} = \sum_{r \in R(t_{kl})} a_{r}$ is any time-independent quantity defined in equations (\ref{Appen:SEgamma}) - (\ref{Appen:SEnu}). Note that $B(t_{k1}), \ldots, B(t_{kq_k})$ can be computed in $O(n)$ operations as one scans through $t_{k1}, \ldots, t_{kq_k}$ backward in time, following the recursive formula (\ref{Appen:MstepScan}). Furthermore, analogous to (\ref{eqn:Lscan}), the following linear scan algorithm can be used to calculate $\{B(T_{(1)}), B(T_{(2)}), \cdots, B(T_{(n)})\}$ from $\{B(t_{k1}),\ldots, B(t_{kq_k})\}$:
\begin{eqnarray*}
\{B(t_{k1}),\ldots, B(t_{kq_k})\} \mapsto \{B(T_{(1)}), B(T_{(2)}), \cdots, B(T_{(n)})\},
\end{eqnarray*}
where for each $t_{kl}$, only a subset of the ranked observation times $T_{(i)}$'s are scanned forward to calculate  $B(T_{(i)})$'s as follows
\begin{equation*}
  B(T_{(i)}) =
    \begin{cases}
      B(t_{k1}), & \text{if $T_{(i)} \geq t_{k1}$,}\\
      B(t_{k(l+1)}), & \text{if $T_{(i)} \in [t_{k(l+1)}, t_{kl}), \; $ for some $l\in\{1,\ldots, q_k-1 \}$,}\\
      0, & \text{otherwise}.
    \end{cases}       
\end{equation*}
Consequently, calculating all $B(T_{(i)})$'s takes $O(n)$ operations.

\end{document}